\documentclass[11pt,a4paper]{article}
\usepackage[centertags]{amsmath}
\usepackage{amsfonts}
\usepackage{amssymb}
\usepackage{amsthm}
\usepackage{graphicx}
\usepackage{ccaption}

\captionnamefont{\bfseries}
  \captiontitlefont{\small\sffamily}
  \captiondelim{: }
  \hangcaption

\author{John Hammersley\footnote{J.C.Hammersley@dur.ac.uk} \\ \\ \textit{Department of Mathematical Sciences,} \\ \textit{Durham University,
South Road, Durham DH1 3LE UK}}
\title{\vspace{-1cm} \begin{flushright} \footnotesize{DCPT-07/15}
\end{flushright} \vspace{1cm} Numerical metric extraction in AdS/CFT}
\date{}
\begin{document}

\maketitle

\begin{abstract}
An iterative method for recovering the bulk information in
asymptotically AdS spacetimes is presented. We consider zero energy
spacelike geodesics and their relation to the entanglement entropy
in three dimensions to determine the metric in certain symmetric
cases. A number of comparisons are made with an alternative
extraction method presented in arXiv:hep-th/0609202, and the two
methods are then combined to allow metric recovery in the most
general type of static, spherically symmetric setups. We conclude by
extracting the mass and density profiles for a toy model example of
a gas of radiation in (2+1)-dimensional AdS.
\end{abstract}

\section{Introduction}

One of the original applications of the holographic principle was in
relating the entropy of a black hole to the area of its horizon
\cite{thooft,suss}; since then a variety of authors have continued
to explore relationships between bulk and boundary physics via
holography, most notably via the AdS/CFT correspondence of Maldacena
\cite{adscft}. The idea of entropy being linked with an area rather
than a volume (as one naturally expects from thermodynamics) is not,
however, restricted to the case of black holes.

Recently, a proposal was put forward by Ryu and Takayanagi
\cite{taka1,taka2} relating the entanglement entropy of a subsystem
in a CFT to the area of a minimal surface in the bulk. This has been
investigated further in a number of subsequent papers, such as
\cite{furs,solo,taka3,hubtak} where a number of related issues are
explored. One avenue of interest leading from this proposal is the
question of whether we can take this link between the entanglement
entropy and minimal surface area, and devise a method to efficiently
extract the bulk physics from the field theory information.

In (2+1) dimensions, the area of the minimal surface in question
corresponds to the length of a static spacelike geodesic connecting
the two endpoints of the region A through the bulk, as illustrated
in figure \ref{stat1}. It is this observation that leads to
comparisons with a method of extracting the bulk metric given in
\cite{hammer}, where the relation between singularities in
correlation functions in the CFT and null geodesics (see \cite{vero}
for details) was used to iteratively recover the bulk metric in
certain asymptotically AdS spacetimes. In this paper we devise a
similar method for extracting the bulk metric, using instead the
relationship of Ryu and Takayanagi between the entanglement entropy
and the length of the relevant spacelike geodesic. Interestingly, we
find that after plotting the proper length against the angular
separation of the endpoints, see figure \ref{lvsphi}, the gradient
$d \mathcal{L} / d \phi$ immediately yields the angular momentum of
the corresponding static spacelike geodesic. This simple relation
then allows the minimum radius of the geodesic to be determined, and
by working iteratively from large $r$, one can reconstruct the
metric function of the bulk.

After describing the method and giving some examples of its
application in practice, we then make a number of comparisons
between this and the method of \cite{hammer} (which is briefly
reviewed in section \ref{sec:review}). Most crucially, the two
methods involve different ways of probing the bulk (as they involve
different types of geodesic path), and whilst they appear
computationally quite similar, this difference allows the method
presented here to probe more fully a greater range of asymptotically
AdS spacetimes. This is a consequence of the fact that in singular
spacetimes, and those with a significant deviation from pure AdS,
the effective potential for the null paths can become non-monotonic,
resulting in geodesics which go into unstable orbits, see figure
\ref{Mv1}. This local maximum in the potential results in a finite
range of radii which cannot be effectively probed by the null
geodesics, and information about the bulk cannot be extracted; one
does not encounter this problem when probing with static spacelike
geodesics, provided the metric function is non-singular. Despite
this advantage, one cannot use either method individually to extract
information from the most general static, spherically symmetric
spacetimes (those with a metric of the form of
\eqref{eq:newAdSmetric}), as neither can provide enough data with
which to fully determine the metric; the null geodesics are not
sensitive to the overall conformal factor of the metric, and the
static spacelike geodesics cannot probe the timelike part. One can,
however, use them in conjunction in order to do so. We thus conclude
by proposing a combination of the two approaches such that the bulk
information can be recovered, and give firstly an example
demonstrating the ease with which it can be done, followed by a toy
model setup of a gas of radiation (a ``star'') in $AdS_{3}$. We
demonstrate how it is possible to determine both the star's mass and
density profiles from our estimates of the metric functions.

The outline of the paper is as follows: Section \ref{sec:backg}
contains background material on asymptotically AdS spacetimes and
geodesic paths, and introduces the entanglement entropy relation
from \cite{taka1}. Section \ref{sec:method1} develops the method for
iteratively extracting the bulk metric, the full details of which
are given in Appendix A, comments on the validity of the solutions,
and goes on to give examples. In Section \ref{sec:compari}, after a
review of the null geodesic approach from \cite{hammer}, the
comparison between this and the spacelike method developed here
follows, where we analyse their similarities and differences in
applicability and efficiency. Finally, the two methods are combined
in Section \ref{sec:combo}, to produce a more generally applicable
method (as illustrated with the recovery of the pertinent
information about a ``star'' in $AdS_{3}$) and we go on to look at
extensions of the method to less symmetric cases in section
\ref{sec:extension}. We conclude in Section \ref{sec:conc} with a
discussion and summary of the results.

\section{Background} \label{sec:backg}

\begin{figure}
\begin{center}
  \includegraphics[width=0.43\textwidth]{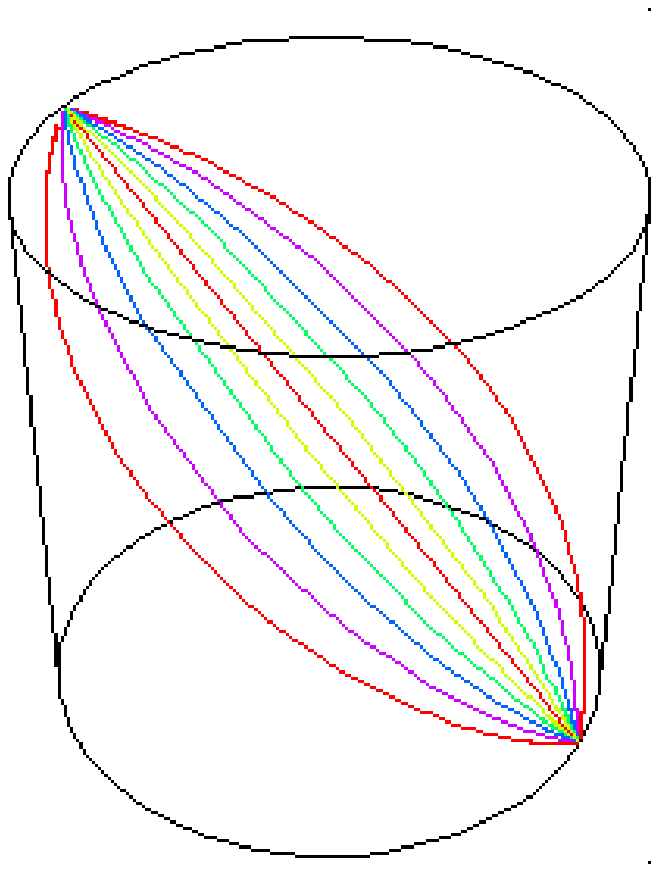}
  \includegraphics[width=0.5\textwidth]{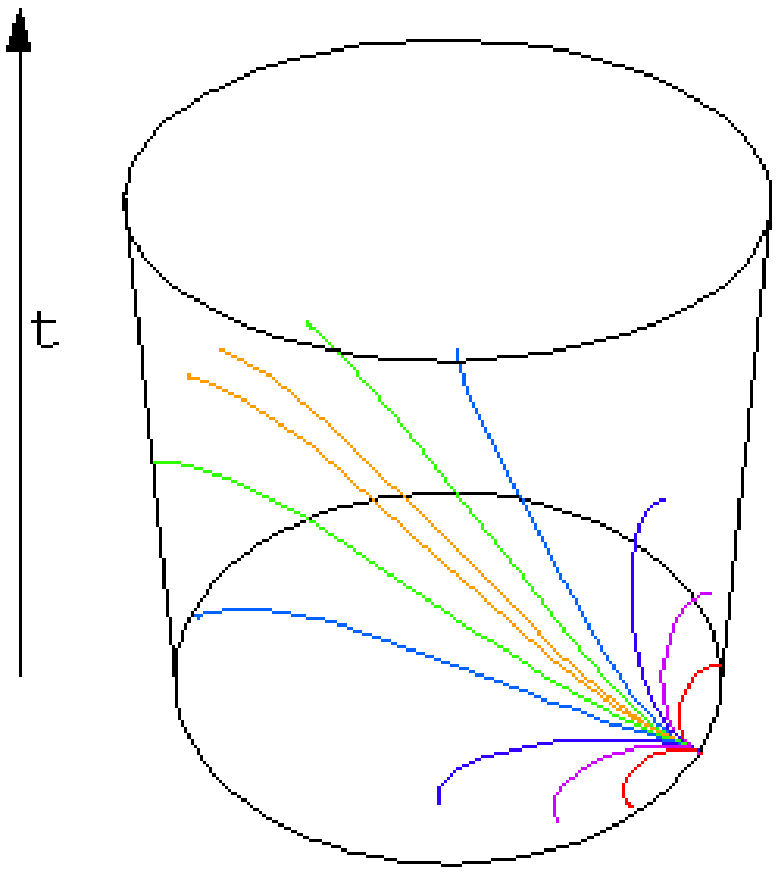}\\
\end{center}
\caption{A sample of geodesic paths in $AdS_{3}$ (with $R = 1$), all
beginning at the same point on the boundary, with varying $J$ and
$E$. The null geodesics (left plot) all terminate at the same
(antipodal) point, whereas this is not the case for spacelike
geodesics (right plot).}\label{geopaths1}
\end{figure}

Recall the metric for $AdS_{3}$ in coordinates $(t,r,\phi)$:
\begin{equation} \label{eq:AdSmetric}
ds^{2} = - f(r) dt^{2} + \frac{dr^{2}}{f(r)} + r^{2}d\phi^{2}
\end{equation}
\begin{equation} \label{eq:f1}
f(r) = 1 + \frac{r^{2}}{R^{2}}
\end{equation}
where $R$ is the AdS radius. The existence of Killing vectors
$\partial/\partial t$ and $\partial/\partial \phi$ leads to two
conserved quantities (energy (E) and angular momentum
(J)\footnote{Note that in \cite{hammer}, the geodesic angular
momentum was denoted $L$; here we use $J$ to avoid confusion with
$L_{T}$, which denotes the length of the system in the CFT (see
section \ref{sec:method1}).}), and allows the geodesic equations to
be written in the simple form:

\begin{equation} \label{eq:AdSveff2}
\dot r^{2} + V_{eff} = 0
\end{equation}
where $\dot \space = \frac{d}{d\lambda}$ for some affine parameter
$\lambda$, and $V_{eff}$ is an effective potential for the
geodesics, defined by:
\begin{equation} \label{eq:AdSveff3}
V_{eff} = - \left(f(r) \kappa + E^{2} - \frac{f(r)
J^{2}}{r^{2}}\right)
\end{equation}
where $\kappa = +1,-1,0$ for spacelike, timelike and null geodesics
respectively. Note that only null and spacelike geodesics can reach
the boundary at $r = \infty$ in finite coordinate time, and so these
are the geodesics we work with when relating bulk physics to the
boundary. The paths of a sample of null and spacelike geodesics
through $AdS_{3}$ are shown in figure \ref{geopaths1}, where one
observes that the null geodesics all terminate at the antipodal
point on the boundary\footnote{This will not be the case in
spacetimes which deviate from pure $AdS_{3}$, see figure
\ref{modnullfig1} in section \ref{sec:review}.}. This is in contrast
to the spacelike geodesic endpoints, where there is a both an
angular and temporal spread in their distribution, obtained by
varying $J$ and $E$ (except in the $E = 0$ case, which we focus on
here, where the geodesics are all contained in a constant time
slice).

Consider a deformation\footnote{This is not the most general
modification one could consider, however, in the more general case,
one needs both null and spacelike probes to determine the metric,
see section \ref{sec:combo}.} to the pure AdS spacetime by replacing
\eqref{eq:f1} with:

\begin{equation} \label{eq:f1new}
f(r) = 1 + \frac{r^{2}}{R^{2}}- p(r)
\end{equation}
where $p(r)$ is an analytic function which is of comparable
magnitude to $r^{2}$ at small $r$ and tends to zero at large $r$.
Now, in \cite{hammer}, the metric information was extracted by using
the endpoints of null geodesics and their relation to correlation
functions in the field theory. Here we propose to use the endpoints
of static spacelike geodesics in three dimensions, and the relation
between their proper length and the entanglement entropy of a two
dimensional CFT proposed in \cite{taka1} to extract the bulk
information.

\subsection{Entanglement entropy} \label{sec:entangle}

\begin{figure}
\begin{center}
  \includegraphics[width=0.48\textwidth]{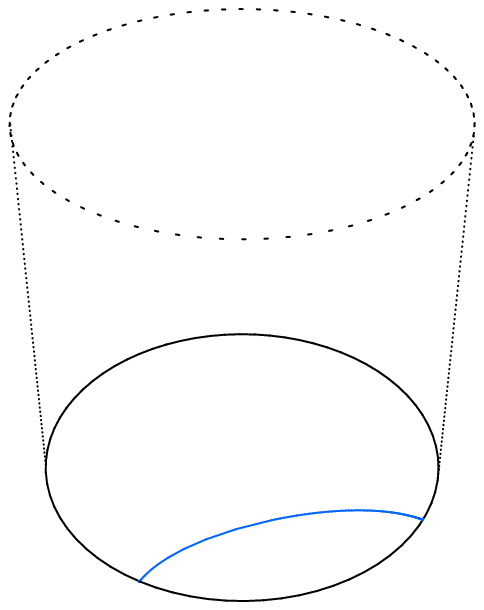}
  \includegraphics[width=0.48\textwidth]{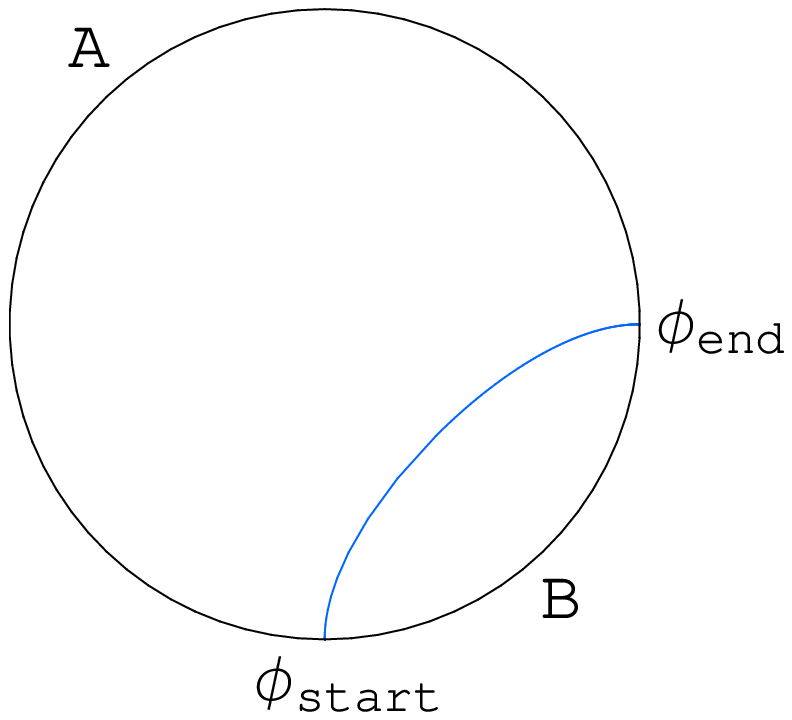}\\
\end{center}
\caption{A static spacelike geodesic in $AdS_{3}$ (left plot), with
the regions A and B highlighted (right plot).}\label{stat1}
\end{figure}

In \cite{taka1}, Ryu and Takayanagi propose that the entanglement
entropy $S_{A}$ (in a $\textrm{CFT}_{d+1}$) of subsystem A with
$(d-1)$-dimensional boundary $\partial A$ is given by the area law:

\begin{equation} \label{eq:entrop1}
S_{A} = \frac{\textrm{Area of }\gamma_{A}}{4 \, G_{N}^{(d+2)}}
\end{equation}
where $\gamma_{A}$ is the static minimal surface whose boundary is
given by $\partial A$, and $G_{N}^{(d+2)}$ is the Newton constant in
$(d+2)$ dimensions. In the $d = 1$ case, $\gamma_{A}$ will be given
by a geodesic line, and thus if we consider $AdS_{3}$ with a
(1+1)-dimensional CFT living on its boundary, and define two regions
A and B on the boundary as in figure \ref{stat1}, Ryu and
Takayanagi's proposal relates the proper length of the static
spacelike geodesic shown to the entanglement entropy $S_{A}$. Thus
by considering a complete set of these geodesics, we can probe the
entire spacetime from out near the boundary down to the centre at $r
= 0$\footnote{This assumes we are working in a non-singular
spacetime; for the case where the central disturbance $p(r)$
corresponds to that for a black hole, one can probe down to the
horizon radius, $r_{h}$, see section \ref{sec:singspaces}.}, as we
discuss in the following section.

\section{Method for reconstructing $f(r)$} \label{sec:method1}

Focussing on spacelike geodesics, and specifically those with zero
energy (i.e. static), we have that:

\begin{equation} \label{eq:zero1}
\dot r^{2} - f(r) \left(1 - \frac{J^{2}}{r^{2}}\right) = 0
\end{equation}
which can be combined with the angular momentum conservation
equation $J = r^{2} \dot \phi$ to give:

\begin{equation} \label{eq:zero2}
\frac{d r}{d \phi} = r \sqrt{f(r)} \sqrt{\frac{r^{2}}{J^{2}} - 1}
\end{equation}
This can then be re-cast as an integral equation along the geodesic
path, where we note that the final angular separation will be a
function of $J$ only:
\begin{equation} \label{eq:phir}
\phi(J) \equiv \int_{\phi_{start}}^{\phi_{end}}\, \mathrm{d} \phi =
2 \int_{r_{min}}^{r_{max}} \frac{1}{r \sqrt{f(r)}
\sqrt{\frac{r^{2}}{J^{2}} - 1}} \, \mathrm{d} r
\end{equation}
where $r_{min}$ is minimum radius obtained by the geodesic, and in
the zero energy case is given simply by $r_{min} = J$. As the metric
is divergent at the boundary $r = \infty$, we introduce a cut-off
$r_{max}$ and restrict ourselves to the region $r <
r_{max}$.\footnote{This cut-off corresponds to the ratio between the
UV cutoff (or equivalently the lattice spacing) in the CFT and the
total length of the system: $r_{max} \sim L_{T}/a$} We also have
that the proper length of the geodesic (also dependent only on $J$)
is given by:

\begin{equation} \label{eq:lengthr}
\mathcal{L}(J) = 2 \int_{r_{min}}^{r_{max}} \frac{1}{\sqrt{f(r)}
\sqrt{1 - \frac{J^{2}}{r^{2}}}} \, \mathrm{d} r
\end{equation}

These two equations, \eqref{eq:phir} and \eqref{eq:lengthr}, will
form the basis for our method of extracting the metric function
$f(r)$ at each $r$.

Now, given that the spacetime in which we are working is
asymptotically AdS, we can say that for $r \ge r_{n}$ for some
$r_{n}$ which can be arbitrarily large (but still below the cut-off
$r_{max}$), $f(r) \approx r^{2} + 1$ (with R set to one). Thus all
static spacelike geodesics with angular momentum $J \ge J_{n} \equiv
r_{n}$ will remain sufficiently far from the central deformation
$p(r)$ such that they remain undisturbed by its effects, and in the
limiting case $J = r_{n}$ we can write:
\begin{eqnarray} \label{eq:phir2}
\phi_{n} && = 2 \int_{r_{n}}^{r_{max}} \frac{1}{r \sqrt{r^{2} + 1}
\sqrt{\frac{r^{2}}{r_{n}^{2}} - 1}} \, \mathrm{d} r \\ && =
\frac{\pi}{2} - \arctan{\left(\frac{2 r_{n}^{2} + \left(r_{n}^{2} -
1\right)r_{max}^{2}}{2 r_{n} \sqrt{r_{max}^{4} - \left(r_{n}^{2} -
1\right)r_{max}^{2} - r_{n}^{2}}}\right)}
\\ && \approx \frac{\pi}{2} - \arctan{\left(\frac{r_{n}^{2} - 1}{2
r_{n}}\right)} \; \; \; \; \textrm{for} \,\, r_{max} \gg r_{n}
\end{eqnarray}
where $\phi_{n} = \phi_{end} - \phi_{start}$, and is the length of
section B of the boundary in figure \ref{stat1}. Hence from the
$\phi$ endpoints, which are specified by the our choice of region A
in the CFT, we can determine $r_{n}$ and we have that $f(r_{n}) =
r_{n}^{2} + 1$. This will be the starting point for an iterative
method which will recover the metric from $r_{n}$ down to zero (in
the non-singular case).

The naive way in which to now proceed is by taking a slightly
smaller choice of minimum radius, $r_{n-1} < r_{n}$, and splitting
up the relevant integrals in \eqref{eq:phir} and \eqref{eq:lengthr}
into two pieces, one from $r_{n-1}$ to $r_{n}$ and one from $r_{n}$
to $r_{max}$. These integrals could then both be well approximated,
the first by taking a series expansion about the minimum radius
$r_{n-1}$, and the second by approximating the spacetime as pure
AdS, as in \eqref{eq:phir2}. We would thus end up with two
simultaneous equations which could be solved to give $r_{n-1}$ and
$f(r_{n-1})$, and could then proceed in a similar fashion to obtain
the the entire bulk metric, to an arbitrary level of accuracy
determined by our choice of step size in $r$ (which is determined by
our choice of boundary region $\phi_{end} - \phi_{start}$). However,
it turns out there is a significant problem with this setup which
prevents it being applied in practice. Specifically, the iterative
process is unstable, with any errors in the estimates for $r_{n-i}$
and $f(r_{n-i})$ leading to greater errors at the next step. This
results in a rapid divergence of the estimate from the actual
metric, and the iteration quickly breaks down. Whilst improving the
approximations to the various terms in the integral can slightly
improve matters, there is a way of avoiding this unstable setup
(where we solve for the two unknowns simultaneously at each step)
entirely, as we shall now demonstrate.

\subsection{Determining the angular momentum} \label{sec:newsecLphi}

Consider the equations \eqref{eq:phir} and \eqref{eq:lengthr} above;
they both have very similar forms, and there is in fact a strikingly
simple yet powerful relation between the two quantities,
$\mathcal{L}$ and $\phi$. Taking the derivative of both with respect
to $J$, the angular momentum, we have that:

\begin{equation} \label{eq:lengthrdj}
\frac{d \mathcal{L}}{d J} = 2 \int_{r_{min}}^{r_{max}}
\frac{J}{r^{2} \sqrt{f(r)} \left(1 -
\frac{J^{2}}{r^{2}}\right)^{3/2}} \, \mathrm{d} r -
\left(\frac{2}{\sqrt{f(r)} \sqrt{1 -
\frac{J^{2}}{r^{2}}}}\right)\Bigg|_{r = r_{min}} \frac{d r_{min}}{d
J}
\end{equation}
and
\begin{equation} \label{eq:phirdj}
\frac{d \phi}{d J} = 2 \int_{r_{min}}^{r_{max}} \frac{1}{r^{2}
\sqrt{f(r)} \left(1 - \frac{J^{2}}{r^{2}}\right)^{3/2}} \,
\mathrm{d} r - \left(\frac{2 \, J}{r^{2} \sqrt{f(r)} \sqrt{1 -
\frac{J^{2}}{r^{2}}}}\right)\Bigg|_{r = r_{min}} \frac{d r_{min}}{d
J}
\end{equation}
Using the fact that $J = r_{min}$, and noting that the divergent
part of the integral cancels with the divergent second term in each
equation\footnote{It is straightforward to show this, and an
equivalent calculation is given explicitly in the second appendix of
\cite{hammer}.}, we can see that the two equations are identical
upto a factor of $J$, and we therefore have that:

\begin{equation} \label{eq:dldphispace}
\frac{d \mathcal{L}}{d J} = J \, \frac{d \phi}{d J}
\end{equation}
which can be rewritten as

\begin{equation} \label{eq:dldphibspace}
\frac{d \mathcal{L}}{d \phi} = J = r_{min}
\end{equation}

\begin{figure}
\begin{center}
  \includegraphics[width=0.9\textwidth]{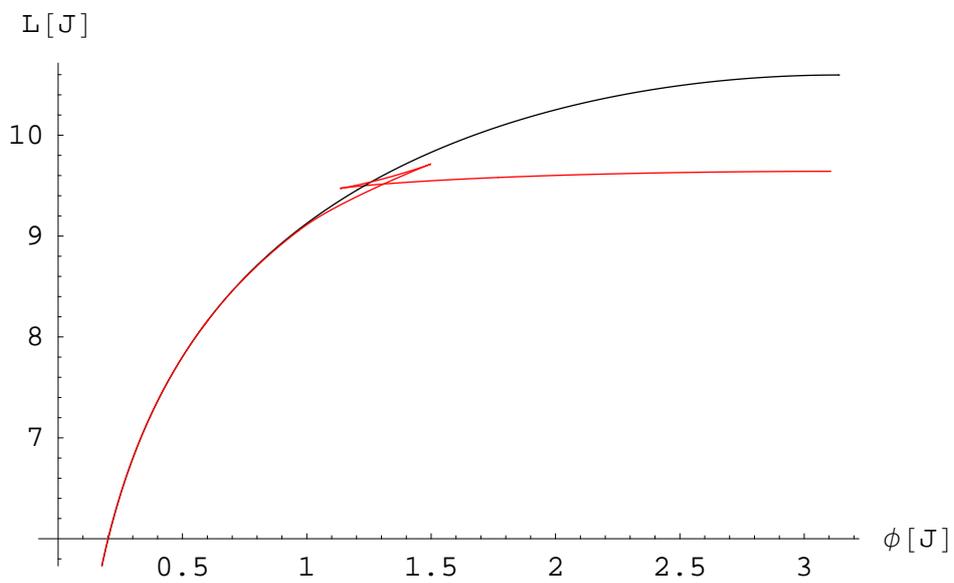}\\
\end{center}
\caption{A plot of the proper length, $\mathcal{L}$, vs the angular
separation of the endpoints, $\phi$, for static spacelike geodesics
in an asymptotically AdS spacetime (red, lower curve), and in pure
AdS (black, upper curve). The gradient, $d \mathcal{L}/d \phi$ at
each point provides the angular momentum, $J$, for the corresponding
geodesic. When the angular separation is small, the geodesics remain
far from the centre, away from the deformation, and hence both
curves coincide.}\label{lvsphi}
\end{figure}

Thus we have the remarkable fact that the minimum
radius\footnote{Note that equation \eqref{eq:dldphibspace} holds in
any static, spherically symmetric spacetime; in those with less
symmetry, such as angular variation of the metric as well as radial,
one finds that the gradient $\frac{d L}{d J}$ gives the final
angular momentum of the geodesic, but as this will not be conserved,
it is not necessarily equal to $r_{min}$.} of the static spacelike
connecting any two points on the boundary is immediately calculable
from the gradient of a plot of the proper length, $\mathcal{L}$
versus angular separation $\phi$, see figure \ref{lvsphi}. This
immediately provides us with one of the two unknowns we need at each
step, and leaves us with only needing to calculate $f(r_{min})$.
This can be done iteratively, beginning at large $r$, by splitting
up \eqref{eq:phir} (or \eqref{eq:lengthr}) and taking various
approximations to each part of the integral, the full details of
which are given in Appendix A. Unlike the original proposal for the
method, this is very robust to any errors, and provides an efficient
way of determining the bulk structure, as we see in the examples in
the following section.

The relation \eqref{eq:dldphibspace} also allows us to more
specifically determine the point at which the metric deviates from
pure AdS; recall that on the first step of the iteration (with
$i=0$), we took the metric to be pure AdS, and after determining
$r_{n}$ using \eqref{eq:phir2}, set $f(r_{n}) = r_{n}^{2} + 1$,
where we originally stated that $r_{n}$ could be taken arbitrarily
large. We can now explicitly check the radii at which the pure AdS
assumption holds, as we can now determine the value of $r_{min}$
corresponding to each $\phi$ separation of the endpoints, and hence
plot $r_{n-i}$ vs $\phi_{n-i}$ for each $i$. In pure AdS, we know
that the relation is given analytically by $r_{min} =
\cot(\frac{\phi_{end} - \phi_{start}}{2})$, and at small enough
angular separation, the two plots should coincide (this is also of
course true on the plot of $\mathcal{L}$ vs $\phi$, see figure
\ref{lvsphi}). This allows one to avoid beginning the iteration at
an excessively large radius, which would reduce the efficiency of
the extraction.

We now address the issue of how confident one can be that the
extracted solution matches the actual metric, before going on to
consider some examples.

\subsection{Validating the extracted solution}
\label{sec:unique}

A natural question to ask at this point is on the uniqueness of the
solution, i.e. is there more than one possible $f(r)$ which gives
the same boundary data for the geodesics? Then if there is a unique
$f(r)$, does this proposal for reconstructing the metric always find
it, and not some alternative set of points $(r_{n-i},f(r_{n-i}))$
which also solve equations \eqref{eq:phir5} and \eqref{eq:lengthr5}
without being the actual metric function?

Considering the second question, it is quite simple to show that if
the metric function $f(r)$ corresponding to the boundary data is
unique, then the iterative method must recover it (up to a level of
accuracy determined by the number of steps). We will show that if
this is not the case, then either the metric function was not
unique, contradicting our assumption, or the estimate does not in
fact correspond properly to the boundary data.

Take the extracted points $(r_{n-i},f(r_{n-i}))$ for $i = 0, \dots
,n $, and use them to construct an interpolation function, which is
then our estimate for the metric function. We can then use this
estimate to compute the proper length and angular separation of all
spacelike geodesics passing through the spacetime. If the generated
data matches with the original data from the field theory, we have
successfully produced an estimate for an actual bulk metric, and by
our assumption of uniqueness, this function must be $f(r)$.

If the generated data fails to match correctly to that from the
field theory, we can deduce that we haven't in fact produced an
estimate for $f(r)$, but instead that our $(r_{n-i},f(r_{n-i}))$ are
simply a set of points which solve the equations \eqref{eq:phir5}
and \eqref{eq:lengthr5}.  In this case, the iterative step size used
to produce the estimate was too large, and the extraction procedure
should be repeated with a smaller step size. Once the new estimate
has been produced, the above test can again be applied; this can
continue until an actual estimate of $f(r)$ is recovered.

Finally, one should note that at an infinitesimally  small step
size, one will use the complete\footnote{By complete, we mean all
geodesics which have minimum radius $r_{min} \le r_{n}$, where
$r_{n}$ can be taken arbitrarily large} set of spacelike geodesics
to probe the spacetime, generating a continuous estimate for $f(r)$
from $r_{n}$ down to zero. As such the data generated from our
estimate must correspond to that from the field theory, as it was
all used in its production. Thus, by uniqueness, the estimate must
correspond to $f(r)$.

A basic argument for the uniqueness of the bulk metric corresponding
to the field theory data (in our case, the proper length of the
static spacelike geodesics as a function of the angular separation
of the endpoints) follows from a comparison of the local degrees of
freedom on each side, by noting that this data and the geometry of
the constant time slice we wish to recreate contain the same amount
of information, as $f(r)$ is a function of the radial coordinate
only. When coupled with the knowledge asymptotic behaviour of the
spacetime (that it approaches pure AdS at large $r$), we have the
boundary conditions needed to ensure that the metric function is
unique. In less symmetric cases one has more freedom in the metric,
but correspondingly one also has more information with which to
determine this, see section \ref{sec:extension} for further comments
on these scenarios.

Finally, one should note that this is simple argument does not
constitute in any way a proof of the existence or uniqueness of the
solution, as here the focus is on demonstrating how an intriguing
link between field theory and the bulk leads to a remarkably simple
process for calculating numerically the corresponding bulk metric.
With this in mind, having argued that with suitable checks the
extracted solution should be an estimate for $f(r)$, we now proceed
to some examples where we examine the accuracy of such estimates.

\subsection{Examples} \label{sec:examples}

To illustrate the procedure for metric extraction, we begin by
considering some examples of deformations of the pure AdS metric. In
the cases considered we have taken the proper length and angular
separation of the endpoints to be known from the relevant field
theory, and taken a linear step size in $J$ (and hence $r_{min}$).
The method of Appendix A is then applied for a variety of step
sizes, and the resulting estimates for $f(r)$ are plotted alongside
the actual curve. The three deviations from pure AdS we consider are
the following:

\begin{equation} \label{eq:fex1}
f_{1}(r) = 1 + r^{2} - \frac{4 \, r^{2}}{(r^{2} + 1)(r^{2} + 8)} +
\frac{3 r \sin(2 \, r)}{r^{4} + 1}
\end{equation}
\begin{equation} \label{eq:fex2}
f_{2}(r) = 1 + r^{2} + \frac{10 \sin^{2}(3 \, r)}{r^{3} + 1}
\end{equation}
\begin{equation} \label{eq:fex3}
f_{3}(r) = 1 + r^{2} + \frac{10 \sin^{2}(10 \, r)}{r^{3} + 1}
\end{equation}
where each gives a non-singular, asymptotically AdS spacetime. These
functions were chosen as tests of the extraction method because they
provide clearly visible deviation from the pure AdS metric of $f(r)
= r^{2} + 1$. The first example also corresponds to one used in
\cite{hammer} in an alternative method for extracting the bulk
information (see section \ref{sec:compari}), and despite the
similarities between $f_{2}(r)$ and $f_{3}(r)$, we shall see a
noticeable difference in the accuracy of their extraction for larger
step sizes.

\begin{figure}
\begin{center}
  \includegraphics[width=0.48\textwidth]{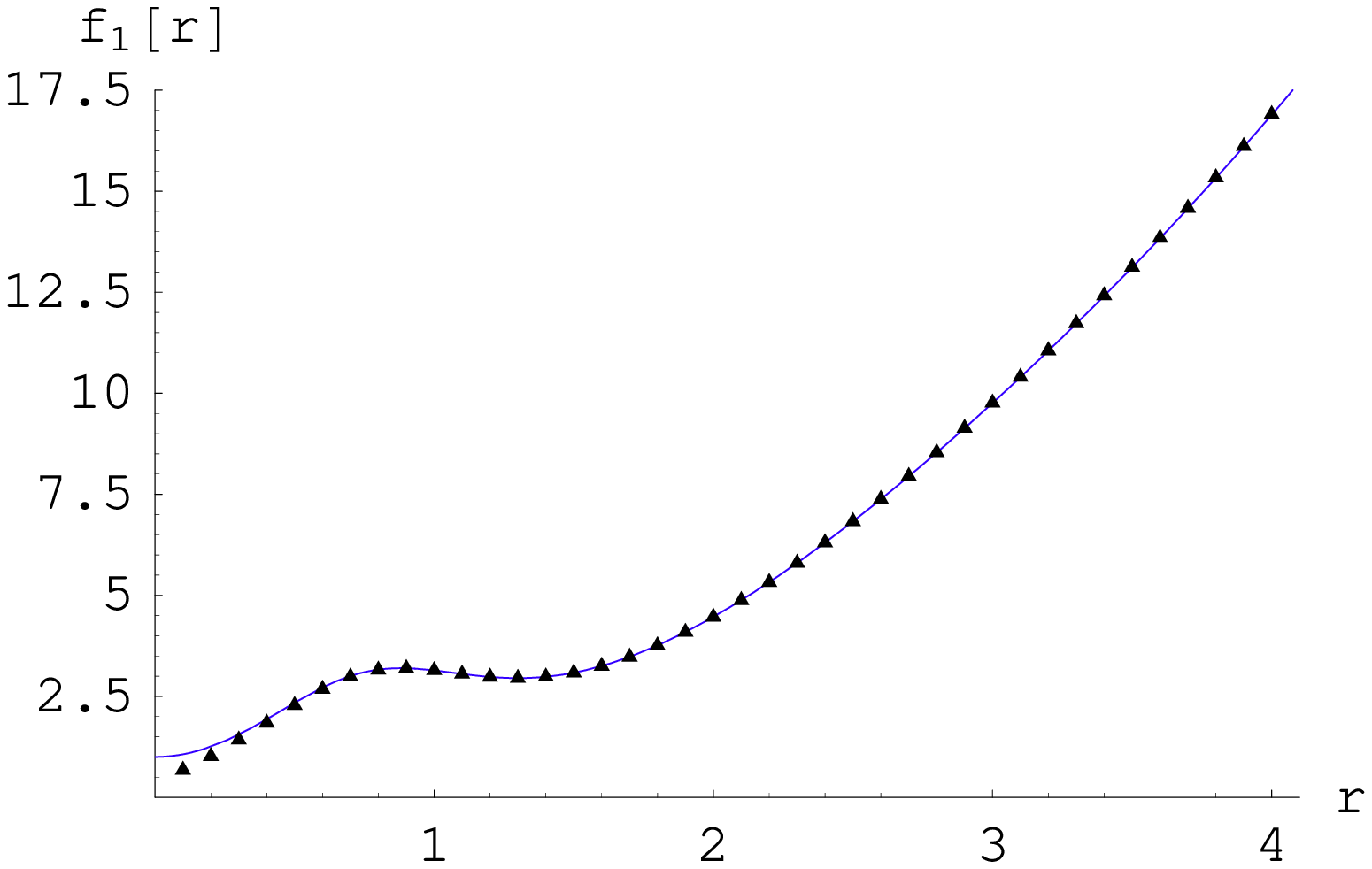}
  \includegraphics[width=0.48\textwidth]{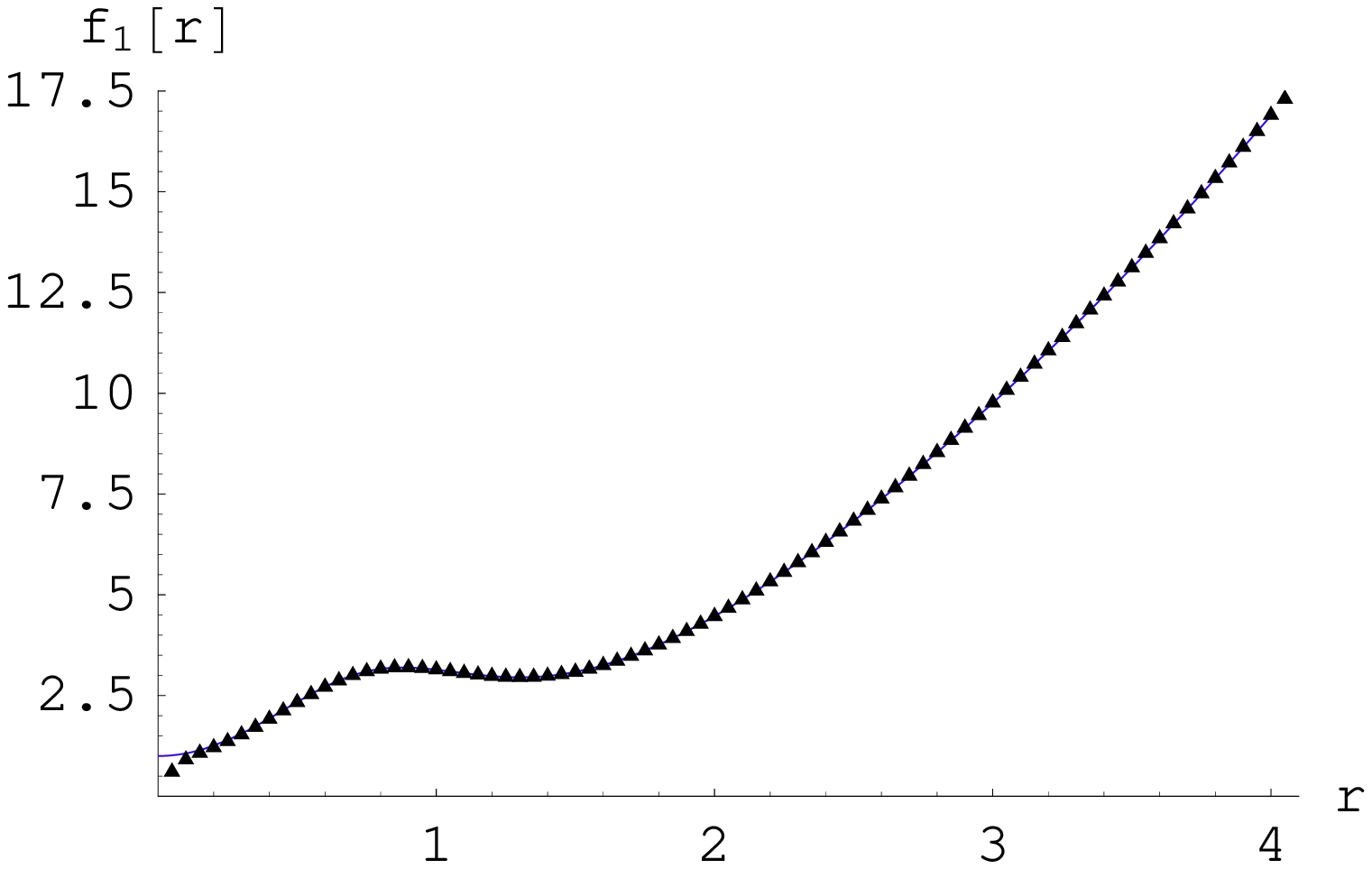}\\
\end{center}
\caption{The data points for the largest two step size estimates for
$f_{1}(r)$, compared with the actual curve (in blue). Whilst both
give good estimates to the curve, the step size of $0.1$ (left)
deviates at a higher $r$ than when using a step size of $0.05$
(right).}\label{f1approx}
\end{figure}

\begin{figure}
\begin{center}
  \includegraphics[width=0.8\textwidth]{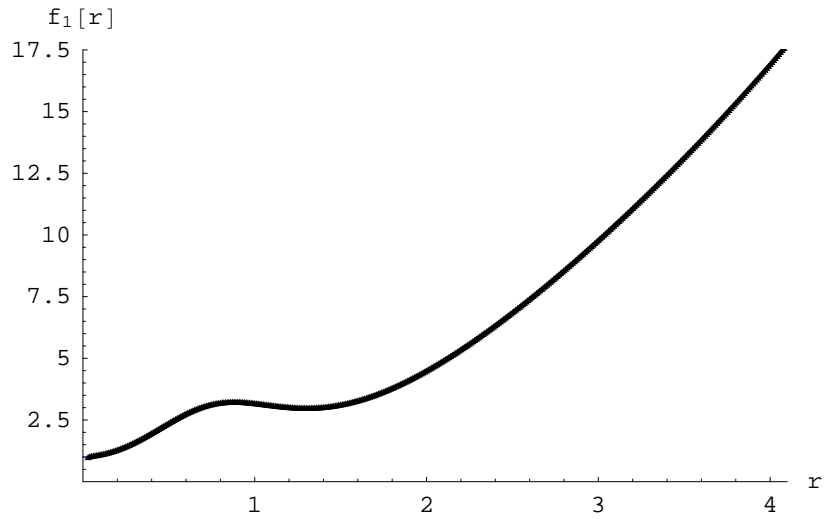}\\
\end{center}
\caption{The data points for the next-to-smallest step size estimate
for $f_{1}(r)$, compared with the actual curve (in blue). The fit
here appears very good even close to $r = 0$, however, Table
\ref{table11} shows that we still need to go to a smaller step size
in order to accurately extract values for $\alpha$, $\beta$ and
$\gamma$.}\label{f1approxb}
\end{figure}

For the first example we use four choices of step size in $r$,
namely $\triangle r \approx 0.1, 0.05, 0.01$ and $0.005$, and
compare the accuracy of the generated curves to the actual function;
this is done by considering best fits to the numerical estimates,
obtained by using a non-linear fit to the following function:

\begin{equation} \label{eq:ffit11}
f_{\textrm{fit} 1}(r) = 1 + r^{2} - \frac{\alpha \, r^{2}}{(r^{2} +
\beta)(r^{2} + \gamma)}+ \frac{\chi r \sin(\eta r)}{r^{4} + \lambda}
\end{equation}
to give values for the various parameters. The results are shown in
Table \ref{table11}, with the corresponding data points plotted in
figures \ref{f1approx} and \ref{f1approxb}.

\begin{table}
\begin{center}
\begin{tabular}{l l l l l l l}
\hline Step size & $\alpha$ (4)& $\beta$ (1)& $\gamma$ (8)& $\chi$ (3)& $\eta$ (2)& $\lambda$ (1)\\
\hline 0.1 & 3.75 & 0.70 & 7.99 & 3.03 & 1.99 & 1.00 \\
0.05 & 3.81 & 0.79 & 7.95 & 3.02 & 1.99 & 1.00 \\
0.01 & 3.94 & 0.85 & 8.19 & 3.01 & 2.00 & 1.00 \\
0.005 & 3.95 & 0.93 & 8.01 & 3.01 & 2.00 & 1.00 \\
\hline
\end{tabular}
\end{center}
\caption{Best fit values (to 2 d.p.) for the $f_{\textrm{fit} 1}(r)$
parameters $\alpha$, $\beta$, $\gamma$, $\chi$, $\eta$ and
$\lambda$, with the actual values indicated in
brackets.}\label{table11}
\end{table}

From Table \ref{table11}, which contains the data for the estimates
of $f_{1}(r)$ we see that there is a very good fit to the actual
values of the six parameters from our non-linear fit
\eqref{eq:ffit11}, even at the largest step size we consider.
Indeed, by eye it is hard to tell any difference between the
accuracy of the estimates except at very small radii. This is mainly
due to the relatively slow variation of $f_{1}(r)$ with $r$, which
ensures the various approximations we take in order to produce the
estimates remain good even for the larger step sizes. Whilst it
appears that taking a smaller step size is rather superfluous, it
should be noted that the finer structure parameters (namely
$\alpha$, $\beta$ and $\gamma$) would need the smaller step size
data in order to be determined to a high level of confidence. Our
choice of non-linear fit function is also rather specifically chosen
to match the example; if one did not know beforehand the form of
$f_{1}(r)$ one would want to take smaller step size estimates in
order to obtain data down as close to $r=0$ as possible (as is
discussed at the end of the section), to ensure that any finer
structure was not being masked, and also as a check on the validity
of the previous estimate.

\begin{figure}
\begin{center}
  \includegraphics[width=0.48\textwidth]{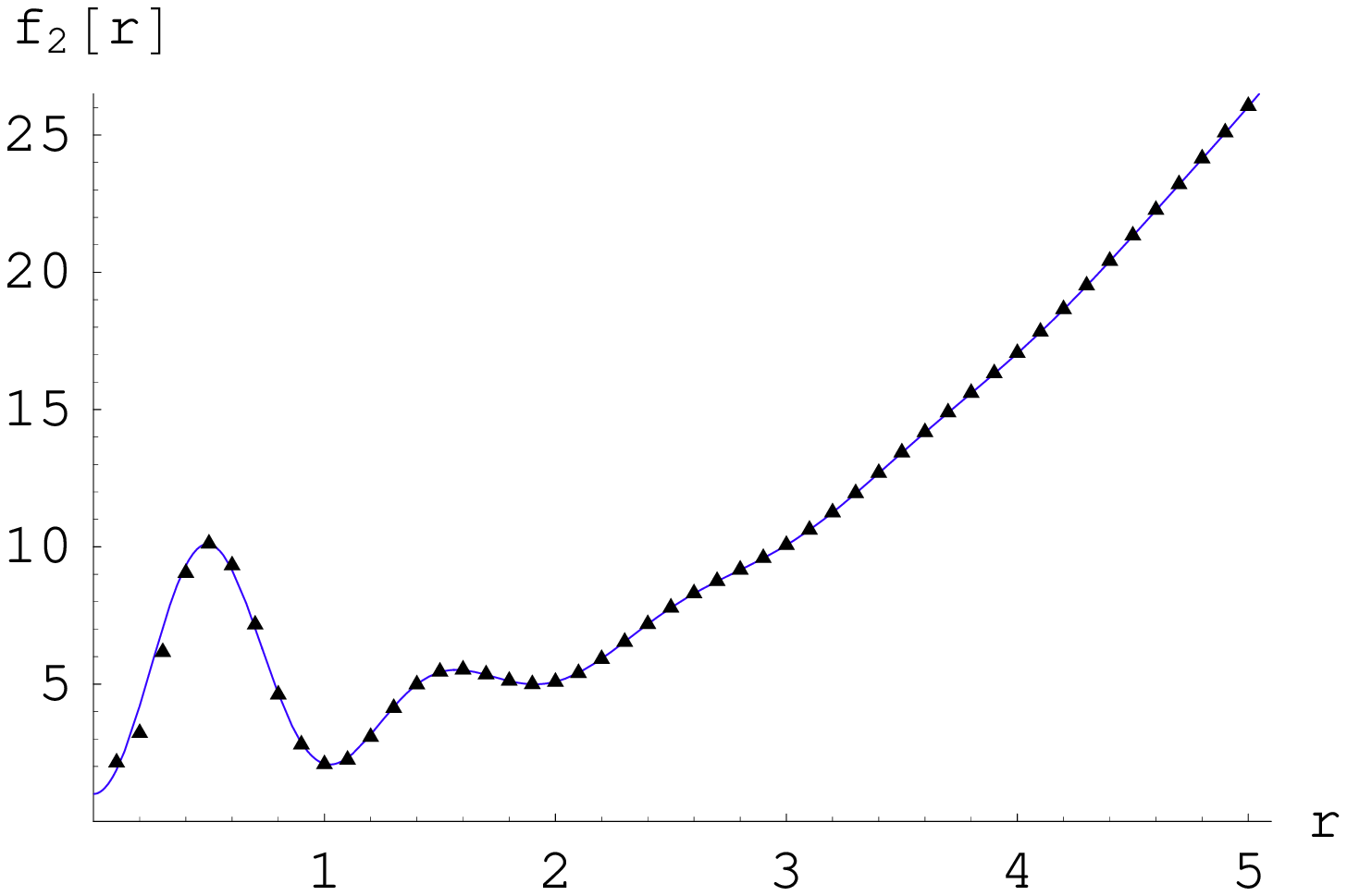}
  \includegraphics[width=0.48\textwidth]{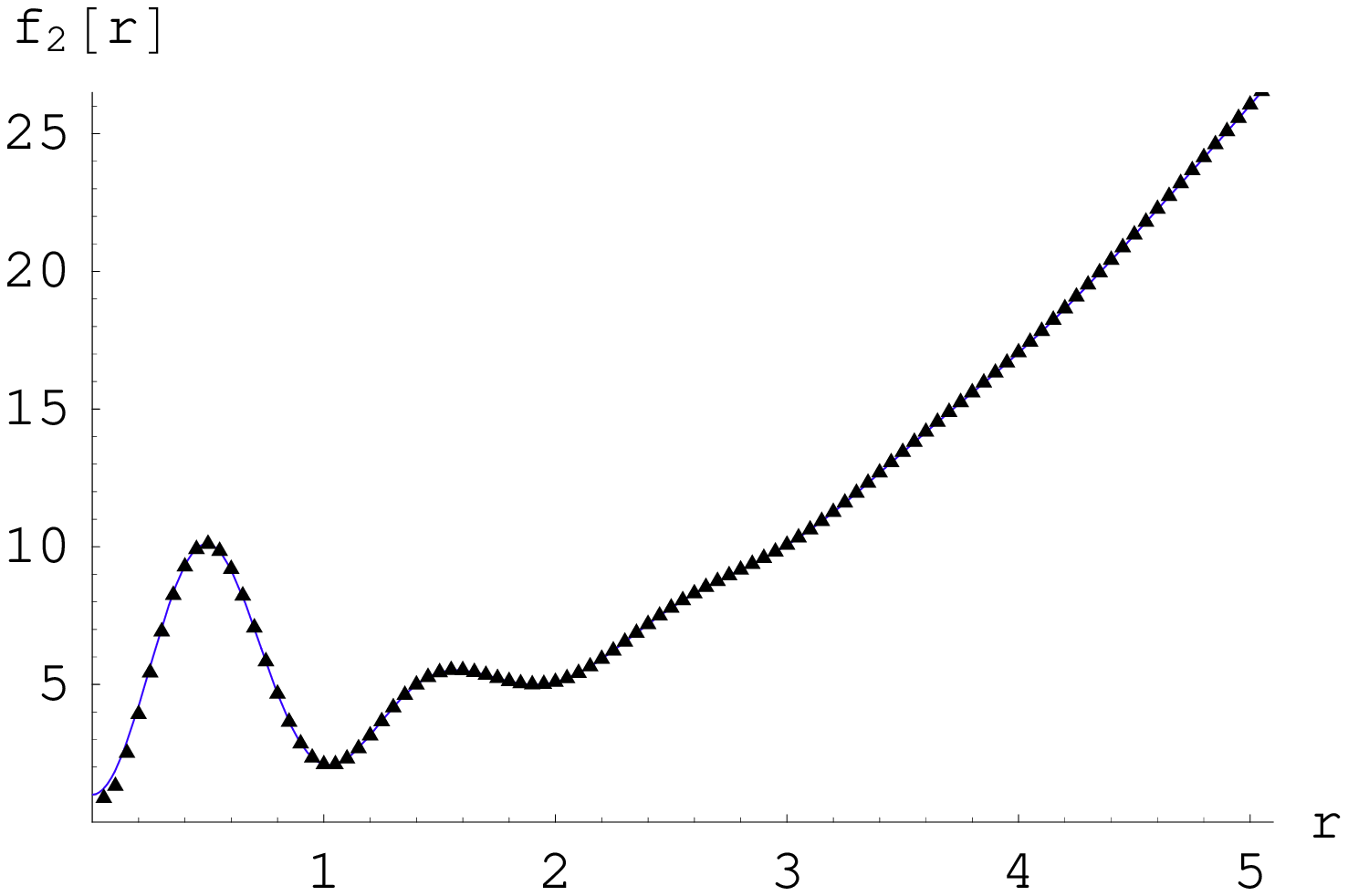}\\
\end{center}
\caption{The data points for the largest two step size estimates for
$f_{2}(r)$, compared with the actual curve (in blue). Despite the
larger deviation from pure AdS than in example 1, both the estimates
here provide good fits to the curve.}\label{f2approx}
\end{figure}

\begin{figure}
\begin{center}
  \includegraphics[width=0.8\textwidth]{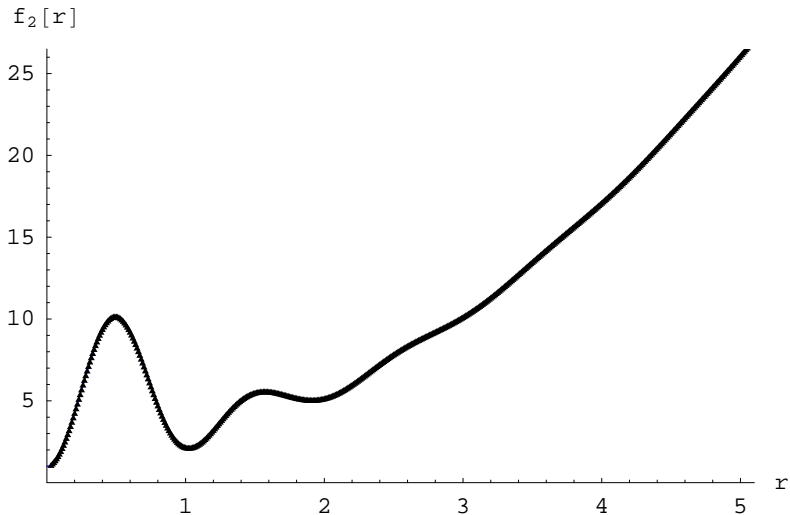}\\
\end{center}
\caption{At a step size of $0.01$, the estimate data for $f_{2}(r)$
matches the actual curve (in blue) almost exactly, even close to $r
= 0$.}\label{f2approxb}
\end{figure}

We see similar behaviour in the second example, where we have chosen
a slightly more fluctuating function to attempt to recover. Here we
use the three largest choices of step size in $r$, and the data
generated in each estimate is shown in figures \ref{f2approx} and
\ref{f2approxb}, where we also include a plot of the actual function
$f_{2}(r)$ as comparison.

\begin{table}
\begin{center}
\begin{tabular}{l l l l}
\hline Step size & $\chi$ (10)& $\eta$ (3)& $\lambda$ (1)\\
\hline 0.1 & 10.32 & 2.99 & 1.06 \\
0.05 & 10.08 & 3.00 & 1.01 \\
0.01 & 10.05 & 3.00 & 1.01 \\
\hline
\end{tabular}
\end{center}
\caption{Best fit values (to 2 d.p.) for the $f_{\textrm{fit} 2}(r)$
parameters $\chi$, $\eta$ and $\lambda$, with the actual values
indicated in brackets.}\label{table22}
\end{table}

We can again use a non-linear fit to evaluate the estimate; in this
case we use a function of the form:
\begin{equation} \label{eq:ffit22}
f_{\textrm{fit} 2}(r) = 1 + r^{2} + \frac{\chi \sin^{2}(\eta \,
r)}{r^{3} + \lambda}
\end{equation}
and the results are shown in Table \ref{table22}.

\begin{figure}
\begin{center}
  \includegraphics[width=0.48\textwidth]{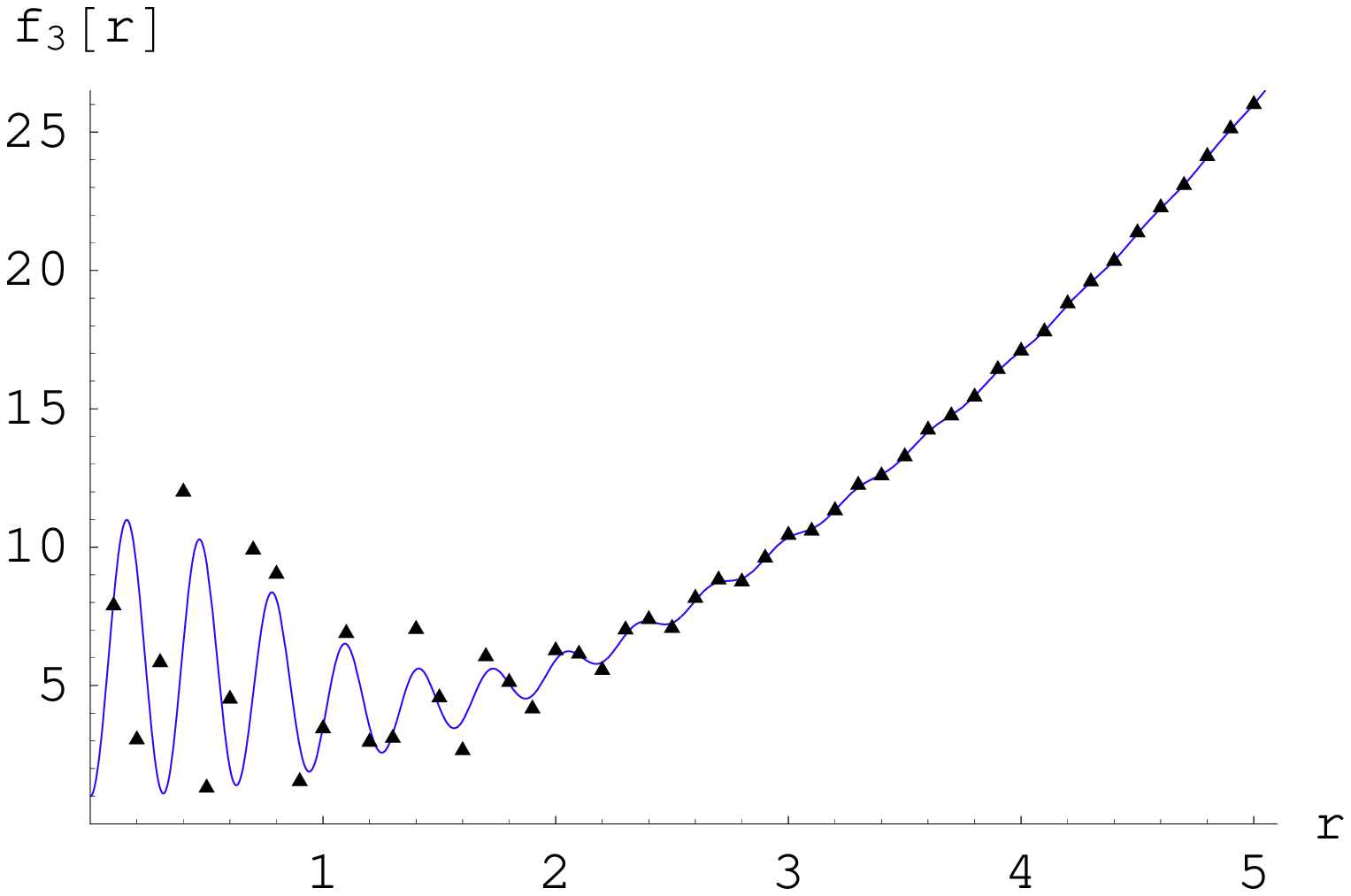}
  \includegraphics[width=0.48\textwidth]{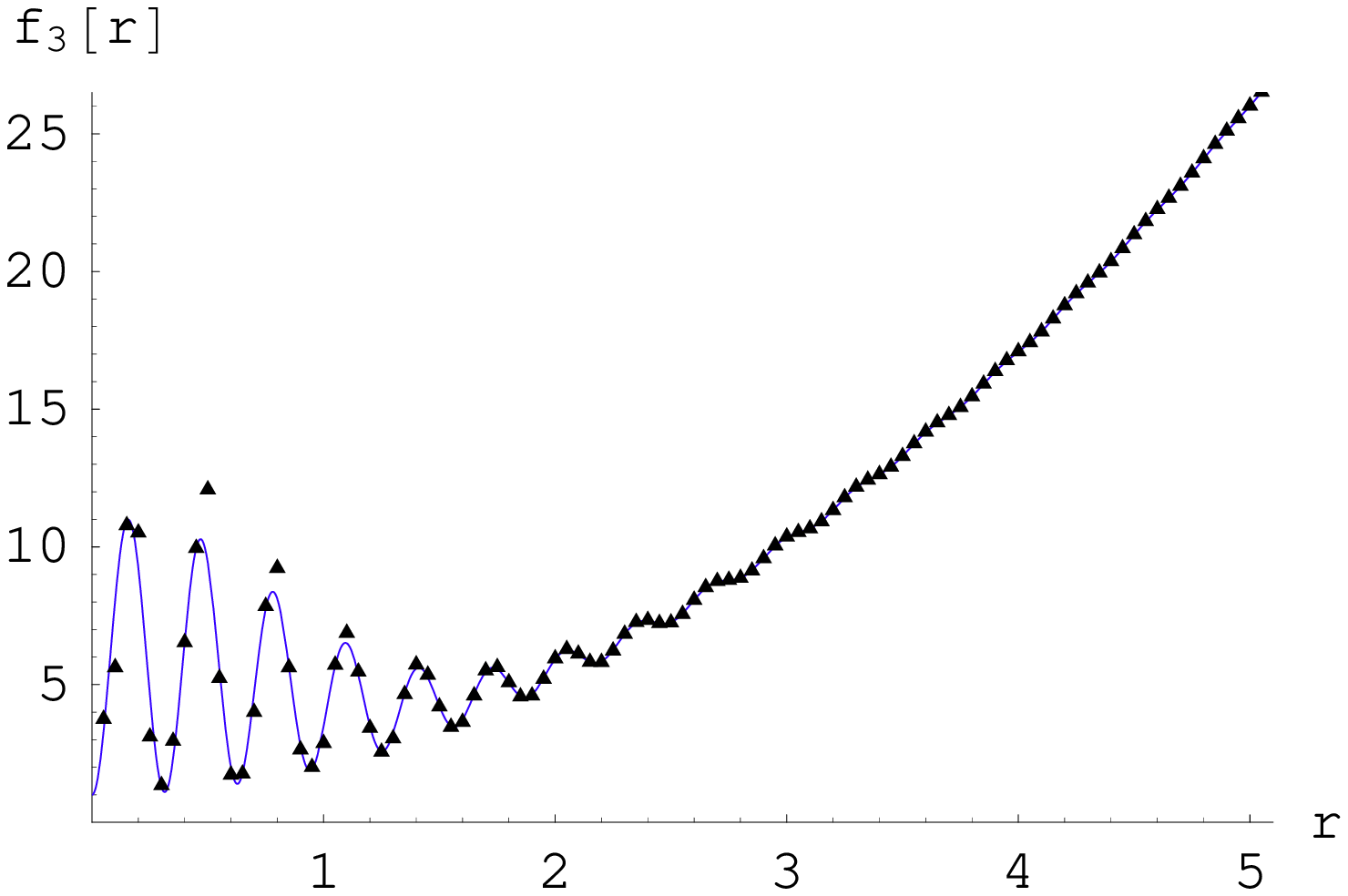}\\
\end{center}
\caption{The data points for the largest two step size estimates for
$f_{3}(r)$, compared with the actual curve (in blue). The reduction
in step size from $0.1$ (left) to $0.05$ (right) gives a marked
improvement in the fit of the points to the curve at low
$r$.}\label{f3approx}
\end{figure}

\begin{figure}
\begin{center}
  \includegraphics[width=0.8\textwidth]{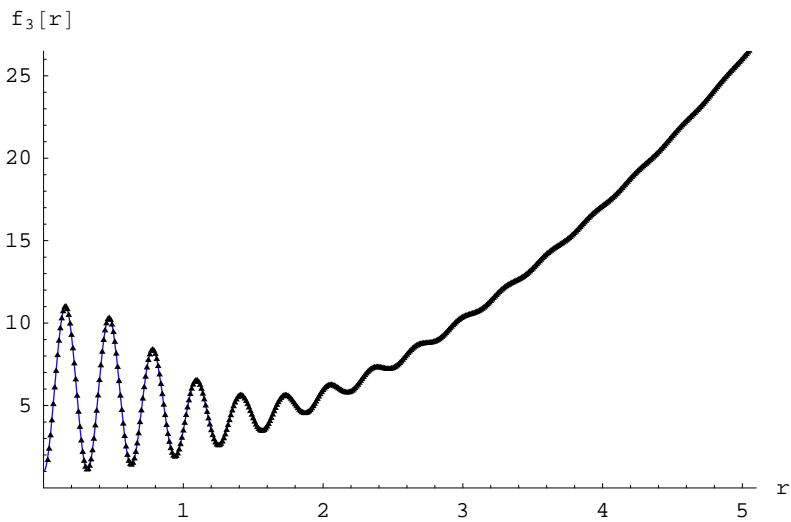}\\
\end{center}
\caption{The data points for the smallest step size estimate for
$f_{3}(r)$, compared with the actual curve (in blue). This level of
precision gives a very good fit to the curve, and this is mirrored
in the highly accurate estimates for the function parameters, given
in Table \ref{table33}}\label{f3approxb}
\end{figure}

Thus far everything is progressing as expected: the smaller step
sizes are producing closer fits to the curve, and better estimates
for the values of the various parameters. In these first two
examples, we even have that the largest step sizes produce good fits
to the curves; do we ever see a large increase in accuracy over our
choice of step size? If we consider the third example (which was
obtained by increasing the value of $\eta$ from the second example),
where the function oscillates more wildly at low $r$, we do see a
significant improvement in the estimates as the step size decreases.
Proceeding as before, we see that for the largest step size of
$0.1$, the method has difficultly in following the rapid
oscillations at low $r$; this is then significantly improved upon in
the subsequent estimates, as shown in figures \ref{f3approx} and
\ref{f3approxb}, and in the non-linear fit data given in Table
\ref{table33}.

\begin{table}
\begin{center}
\begin{tabular}{l l l l}
\hline Step size & $\chi$ (10)& $\eta$ (10)& $\lambda$ (1)\\
\hline 0.1 & 7.49 & 8.03 & 0.29 \\
0.05 & 11.60 & 10.00 & 1.25 \\
0.01 & 9.96 & 9.99 & 0.99 \\
\hline
\end{tabular}
\end{center}
\caption{Best fit values (to 2 d.p.) for the $f_{\textrm{fit} 2}(r)$
parameters $\chi$, $\eta$ and $\lambda$, with the actual values
indicated in brackets.}\label{table33}
\end{table}

As expected, the smaller step size again produces a closer fit to
the actual curve, however, in this third example, the largest step
size fail to give accurate data for the unknowns $\chi$, $\eta$ and
$\lambda$, although it does make a reasonably close fit to the curve
until the iterative process breaks down.

Finally, one should comment on the fact that the deviation of the
estimate from the actual curve does not apparently prevent the
iteration from continuing to give sensible looking (although
erroneous) values in subsequent steps. Whilst appearing to allow for
an incorrect determination of the metric, applying the checks
described in section \ref{sec:unique} (reconstructing the field
theory data using the metric estimate) will quickly highlight any
areas in which the estimate for $f(r)$ has deviated from the correct
function. As stated before, this merely indicates that the step size
in $r$ was too great for the iterative method to properly be
effective in extracting the information using the approximations
chosen in Appendix A. Aside from simply reducing the step size, or
using better approximations (such as at each step creating an
interpolating function estimate for $f(r)$ using the already
determined data), there are other possible resolutions of this
problem to further optimise the extraction. One could take either a
non-linear step size in $r$ to include more terms near $r = 0$, or
simply take appropriately varying step sizes depending on the
fluctuations of the metric; where the metric is varying rapidly with
$r$ the step size could be reduced. Thus by making several passes,
reducing the step sizes at appropriate $r$ each time, the estimate
of $f(r)$ can be significantly improved without considerably
increasing the computation time.

We now conclude the examples section by briefly investigating how
the method is affected in spacetimes with a wildly fluctuating
interior, and how one can apply the above to maintain a high degree
of accuracy.

\subsection{Maintaining accuracy in wildly fluctuating spacetimes}

The third example of the previous section has shown that in wildly
fluctuating spacetimes one needs smaller step sizes in order to
guarantee accuracy of the estimate for $f(r)$ down to small $r$.
Here we provide two further examples to show how the method breaks
down if the frequency of the fluctuations is sufficiently increased,
and how one can adjust the step size to compensate.

\begin{figure}
\begin{center}
  \includegraphics[width=0.48\textwidth]{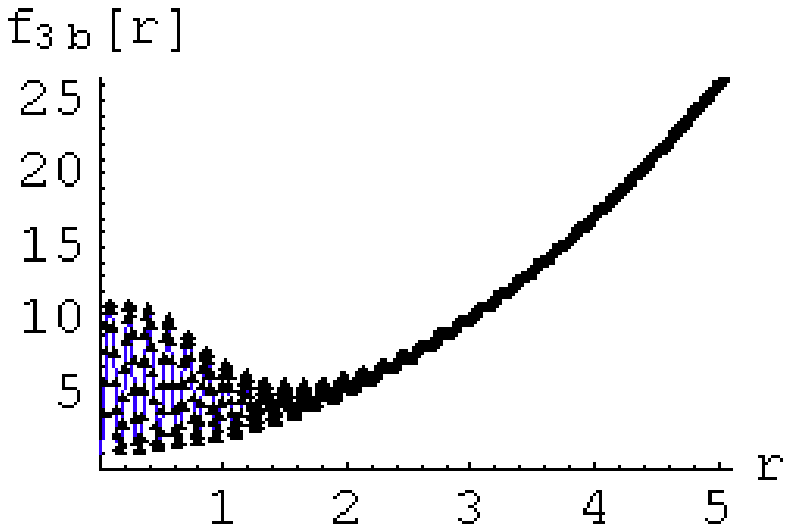}
  \includegraphics[width=0.48\textwidth]{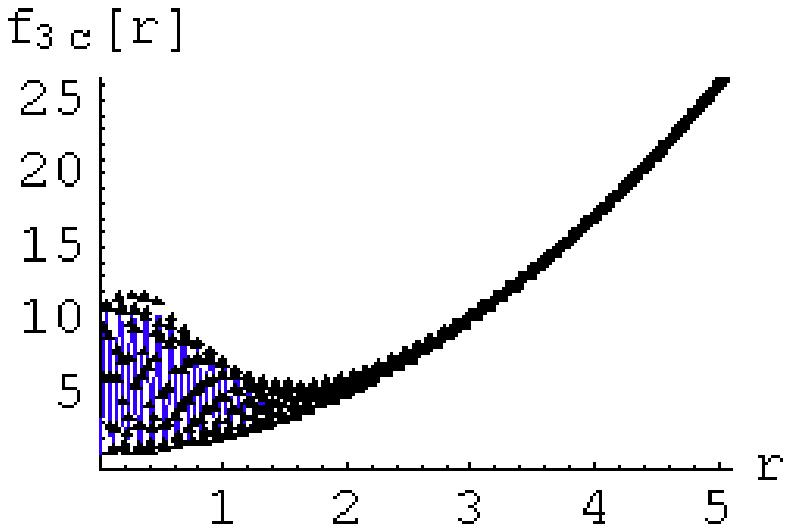}\\
\end{center}
\caption{Plots of $f_{3}(r)$ with the $\sin^{2}(10 \, r)$ term
replaced by $\sin^{2}(20 \, r)$ (left) and $\sin^{2}(30 \, r)$
(right), along with estimates generated with a step size of $0.01$.
Interestingly, whilst sufficiently increasing the frequency of the
metric oscillations does reduce the depth to which the metric is
accurately extracted, it does not adversely affect the accuracy of
the fit to that point.}\label{f3bcapprox}
\end{figure}

Firstly, one observes that it is not simply the frequency of the
oscillation which causes the extraction to break down, but also the
amplitude; this can be seen in figure \ref{f3bcapprox}, where the
estimate continues to follows the actual curve closely whilst the
amplitude of the oscillations is small. The two examples shown in
the figure come from considering modifications to example 3 where
the $\sin^{2}(10 \, r)$ term is replaced by first $\sin^{2}(20 \,
r)$ and then $\sin^{2}(30 \, r)$; as stated, one still obtains a
relatively good fit to the curve using the smallest step size,
although in the more rapidly oscillating case the fit does deviate
slightly more from the correct curve, especially near the peaks at
low $r$.

This behaviour is important, as it means that even in metrics with a
large and rapidly varying interior, one can use a reasonable step
size to extract the metric with confidence down to a fairly close
distance to the centre. After checking the estimate by recreating
the field theory data, one can then continue the extraction from
that point with better approximations, and a smaller step size
(beginning slightly further out than the final terms so as to give
some overlap with the initial estimate and check the consistency of
the estimates) in order to fully reconstruct the metric function.

In any case, the more exotic spacetimes one might wish to consider
may not have only one independent metric function $f(r)$ to extract,
and in order to fully determine the metric in these more general
cases, one may also need to consider the use of null geodesic
probes. Thus having now established the principles of the method,
and seen some examples, we go on to look at comparisons with an
alternative method of extracting the bulk metric proposed in
previous work.

\section{Comparison with an alternative approach to metric extraction} \label{sec:compari}

After seeing in the previous section examples of how the extraction
works in practice, we now consider how this method (S) based on
spacelike geodesics compares to an alternative method involving null
geodesics (N). Before we do so, however, we firstly provide a short
review of this different approach to probing the bulk, which was
originally presented in \cite{hammer}.

\subsection{Review of the null geodesic extraction method} \label{sec:review}

\begin{figure}
\begin{center}
  \includegraphics[width=0.40\textwidth]{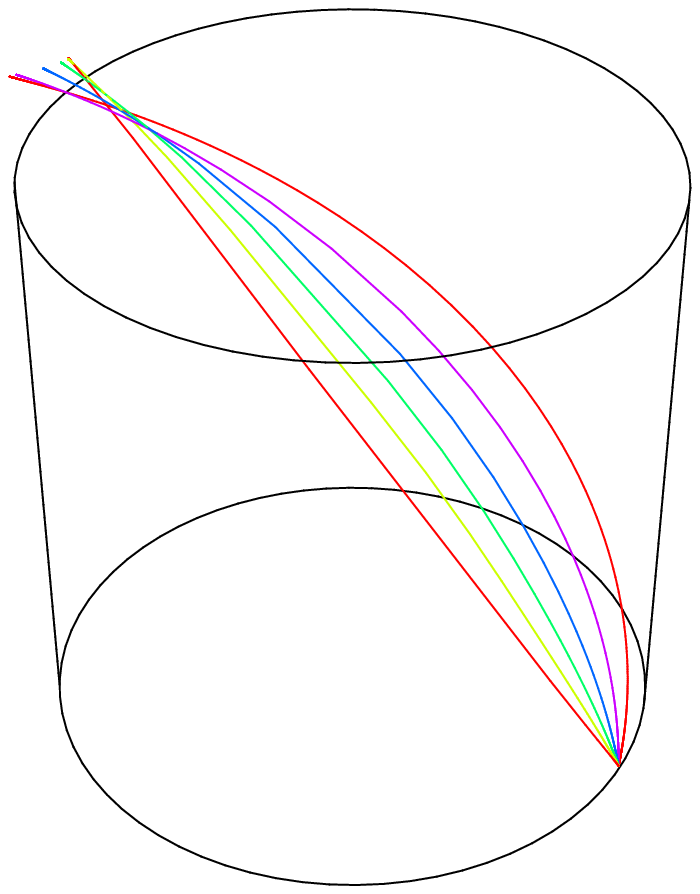}
  \includegraphics[width=0.50\textwidth]{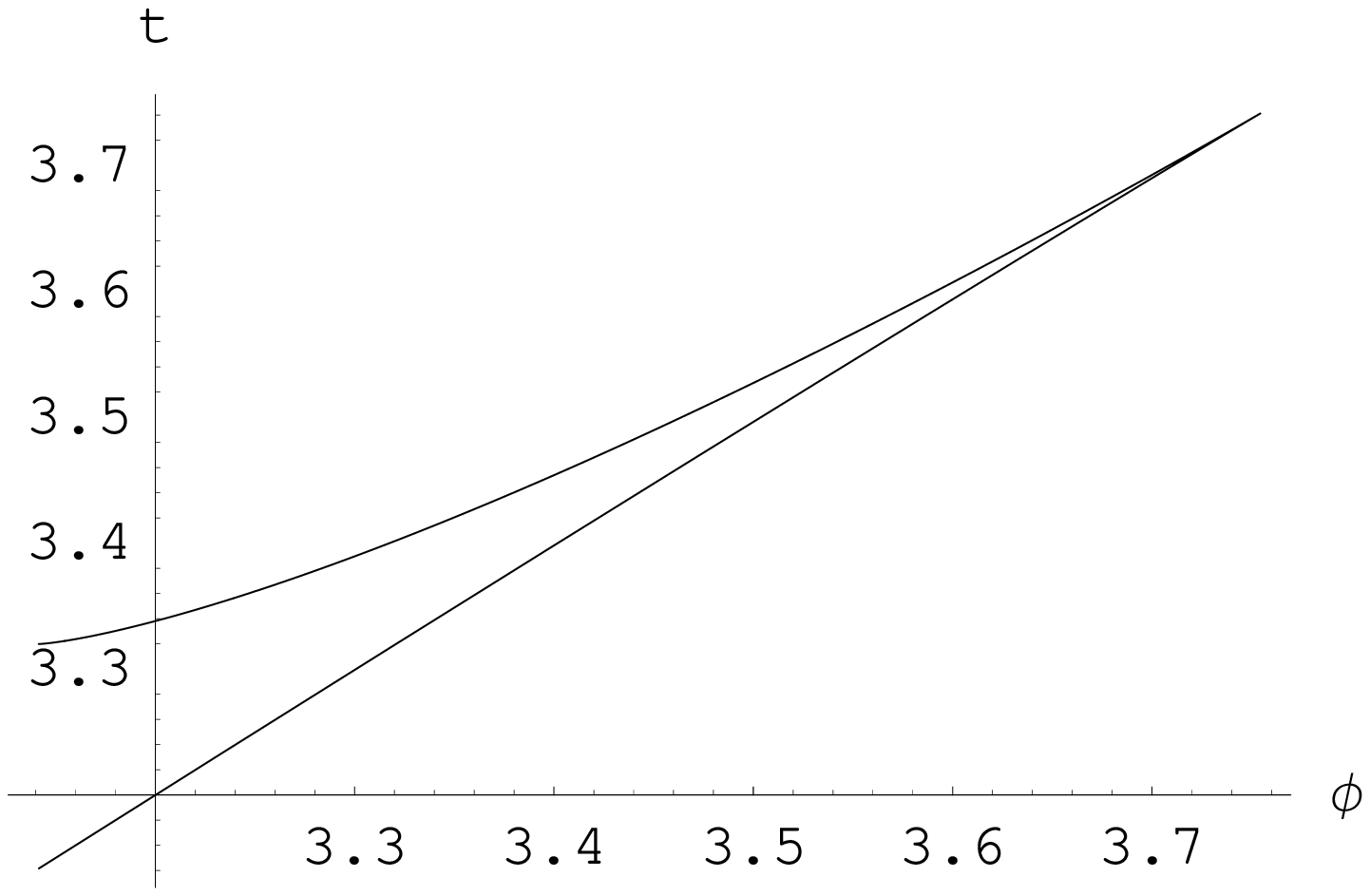}\\
\end{center}
\caption{Null geodesic paths passing through a modified AdS
spacetime, all starting from the arbitrary point $t=0$, $\phi = 0$
on the boundary and with $y >0$. The corresponding full spectrum of
null geodesic endpoints for this spacetime is shown on the right.
(Redrawn figure from \protect\cite{hammer})}\label{modnullfig1}
\end{figure}

For a spacetime of the form of \eqref{eq:AdSmetric} with metric
function $f(r)$ as in \eqref{eq:fex1} say, we can consider the full
set of null geodesic paths through the bulk, which is obtained by
choosing some arbitrary starting point on the boundary and varying
the ratio, $y = J/E$ from zero to one, see figure \ref{modnullfig1}

From this plot of the endpoints, if one takes the gradient
$dt/d\phi$ at any point, one obtains the value of $y$ for the
corresponding geodesic. This is in a sense the ``extra'' piece of
information (analogous to considering $d \mathcal{L}/d\phi$, see
section \ref{sec:newsecLphi}) determined from the CFT which allows
the geodesic probes to extract the bulk metric; here it is the ratio
$y$ of angular momentum to energy which is obtained, in the
spacelike geodesic method it was simply $J$. After determining the
first term of the iteration by taking the spacetime to be pure AdS
far away from the centre, one can then take similar approximations
to those given in Appendix A to split up the relevant geodesic
equation:

\begin{equation} \label{eq:tr}
\int_{t_{0}}^{t_{1}}\, \mathrm{d} t = 2 \int_{r_{min}}^{\infty}
\frac{1}{f(r) \sqrt{1 - y^{2} \frac{f(r)}{r^{2}}}} \, \mathrm{d} r
\end{equation}
and combine with the relation $y = r_{min}/\sqrt{f(r_{min})}$ to
iteratively extract the metric. At this point it is worth making a
computational observation about the two approaches; both involve
almost identical procedures for iteratively extracting the metric,
and as such are of comparable efficiency. There are, however, a
number of fundamental differences between them, as we shall now
discuss.

\subsection{Dimensional applicability} \label{sec:dims}

Whilst in \cite{hammer} method N was applied to the specific case of
$AdS_{5}$, it is equally applicable in an arbitrary dimensional
spacetime, $AdS_{n+1}$ (for $n \ge 2$), assuming one could obtain
the endpoint information from the appropriate field theory on the
boundary. Whilst the principles of method S can also be applied in
arbitrary dimensions, it is no longer clear as to whether the proper
length of the spacelike geodesic is so readily extractable from the
CFT in anything other than the $n = 2$ case. In higher dimensions,
the area of the minimal surface which corresponds to the
entanglement entropy is no longer the length of a spacelike
geodesic, and the method would need to be modified to take this into
account. This could be achieved either by using some expression for
the minimal surface instead of the proper length equation
\eqref{eq:lengthr}, or by demonstrating an alternative route to
determining the proper length.

\subsection{Singular spacetimes and those with significant deviation from pure
AdS} \label{sec:singspaces}

\begin{figure}
\begin{center}
  \includegraphics[width=0.8\textwidth]{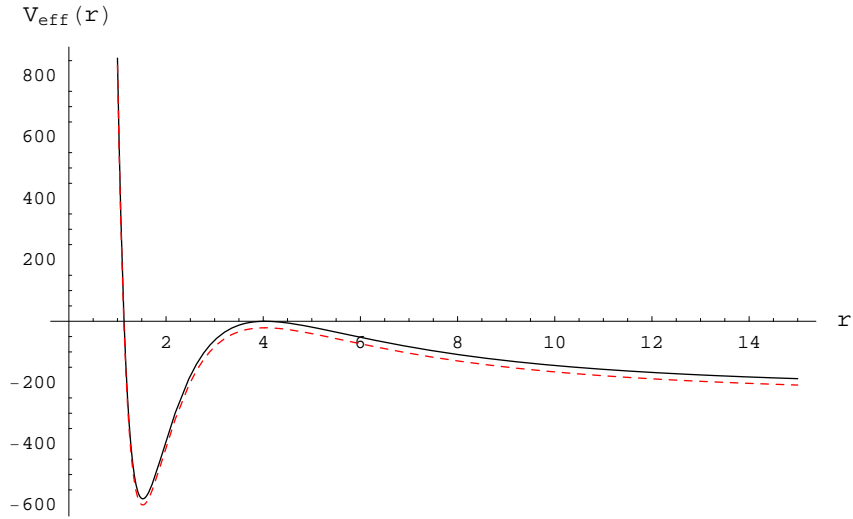}\\
\end{center}
\caption{Plot of the effective potential for two null geodesics with
similar $y$, in some arbitrary spacetime. The null probe which
follows the solid effective potential will go into circular orbit
due to the local maximum; the geodesic with slightly lower $y$
(dashed red line) then has significantly lower $r_{min}$, and this
finite jump in the minimum radius causes the iterative extraction
method to break down.}\label{Mv1}
\end{figure}

One of the main limitations of method N is that it cannot probe past
a local maximum in the effective potential for the null geodesics
(see figure \ref{Mv1}); it cannot therefore probe close to the
horizon of a black hole for instance. The method presented here
would have no such problem, as the spacelike geodesics can reach
arbitrarily close to the horizon while still being able to return to
the boundary. For example, consider a five dimensional
Schwarzschild-AdS spacetime with metric function $f(r)$ given by:

\begin{equation} \label{eq:f3}
f(r) = 1 + r^{2}- \frac{2}{r^{2}}
\end{equation}
where we have set $r_{h} = R = 1$. As was shown in \cite{hammer},
using method N one is only able to probe down to a radius of $r =
2$, as at this point the effective potential for the null geodesics
becomes a local maximum. Method S, however, allows the bulk
information to be fully determined all the way to the horizon
radius, $r_{h} = 1$. Similarly, for those non-singular spacetimes
with large enough deviation from pure AdS so as to allow for null
geodesic orbits (the signature of a non-monotonic effective
potential), one has no problem extracting the full metric using
method S, as in the second and third examples of section
\ref{sec:examples}.

\subsection{The overall conformal factor}

Finally, one should point out that the method presented here is also
sensitive to the overall conformal factor of the metric, whereas
method N is not. This simply stems from the fact that for null
geodesics, $d s^{2}$ is zero, and hence for any metric:

\begin{equation} \label{eq:confAdSmetric}
ds^{2} = \Omega(r) \left( - f(r) dt^{2} + \frac{dr^{2}}{f(r)} +
r^{2}d\phi^{2} \right)
\end{equation}
the conformal factor immediately drops out. For spacelike geodesics
however, $d s^{2} = 1$, and thus the $\Omega(r)$ term remains.
Whilst this conformal factor $\Omega(r)$ presents us with another
unknown, we shall see in the following section how it can be
determined by combining the two methods (N and S) together.

\section{Applying the two methods together} \label{sec:combo}

Having compared the relative merits of the two methods, we now
investigate how it is possible to use them in conjunction with one
another to determine the metric in more general cases. Thus far we
have restricted ourselves to considering metrics of the form of
\eqref{eq:AdSmetric}, however, we can look to extend this further by
considering the most general static, spherically symmetric
spacetimes, given by metrics of the form:

\begin{equation} \label{eq:newAdSmetric}
ds^{2} = - f(r) dt^{2} + h(r) dr^{2} + r^{2}d\phi^{2}
\end{equation}
where we have incorporated the conformal factor $\Omega(r)$ of
\eqref{eq:confAdSmetric} into two new functions $f(r)$ and $h(r)$
(and rescaled the radial coordinate accordingly). Using either
method independently to recover the metric fails because of the
presence of three unknowns: $r$, $f(r)$, and $h(r)$ with only two
independent equations with which to determine them. We can, however,
use both methods in conjunction, as outlined below, where we
restrict ourselves to the (2+1)-dimensional case in accordance with
section \ref{sec:dims}.

For a spacetime of the form of \eqref{eq:newAdSmetric}, we have the
two constraints on the energy and angular momentum from before:
\begin{equation} \label{eq:AdSconstraint1new}
E = f(r) \dot t
\end{equation}
\begin{equation} \label{eq:AdSconstraint2new}
J = r^{2} \dot \phi
\end{equation}
along with the modified expression involving the effective
potential:
\begin{equation} \label{eq:AdSveffc2new}
\dot r^{2} - \left(\frac{\kappa}{h(r)} + \frac{E^{2}}{f(r) h(r)} -
\frac{J^{2}}{h(r) r^{2}}\right) = 0
\end{equation}
We immediately see that for the zero energy spacelike geodesic paths
we do not obtain any information about the function $f(r)$ (as we
would expect, as $f(r)$ affects the time coordinate, which is kept
constant in the $E = 0$ case), and our integrals for the separation
of the endpoints and proper length are given by:

\begin{equation} \label{eq:phirnew}
\phi_{end} - \phi_{start} = 2 \int_{r_{min}}^{r_{max}}
\frac{\sqrt{h(r)}}{r \sqrt{\frac{r^{2}}{J^{2}} - 1}} \, \mathrm{d} r
\end{equation}
and
\begin{equation} \label{eq:lengthrnew}
\mathcal{L} = 2 \int_{r_{min}}^{r_{max}} \frac{\sqrt{h(r)}}{ \sqrt{1
- \frac{J^{2}}{r^{2}}}} \, \mathrm{d} r
\end{equation}

We can thus use the static spacelike geodesics to determine $h(r)$,
from $r = 0$ to an arbitrarily large $r_{n}$, by applying the
extraction method proposed in section \ref{sec:method1} and Appendix
A. Specifically, for each $r_{i}$ we have the corresponding
$h(r_{i})$, and from this one can generate a best fit curve,
$h_{\textrm{fit}}(r)$. One then is left with extracting the $f(r)$
information from the null geodesic endpoints: for a null geodesic in
a bulk with metric \eqref{eq:newAdSmetric}, we have that

\begin{equation} \label{eq:trcnew}
\int_{t_{start}}^{t_{end}}\, \mathrm{d} t = 2
\int_{r_{min}}^{\infty} \frac{\sqrt{h(r)}}{f(r) \sqrt{\frac{1}{f(r)}
- \frac{y^{2}}{r^{2}}}} \, \mathrm{d} r
\end{equation}
with the minimum radius given by $y = r_{min}/\sqrt{f(r_{min})}$. If
we now replace the function $h(r)$ with our estimate
$h_{\textrm{fit}}(r)$, this becomes
\begin{equation} \label{eq:trcnew2}
\int_{t_{start}}^{t_{end}}\, \mathrm{d} t = 2
\int_{r_{min}}^{\infty} \frac{\sqrt{h_{\textrm{fit}}(r)}}{f(r)
\sqrt{\frac{1}{f(r)} - \frac{y^{2}}{r^{2}}}} \, \mathrm{d} r
\end{equation}
which contains only two unknowns, as the parameter $y$ is given by
the gradient of the endpoints (see section \ref{sec:review}). We can
then use the iterative method of \cite{hammer} (the relevant
equations are given in Appendix C) to recover the second metric
function, $f(r)$, and the bulk information has been extracted, as we
see for the two examples which follow. The main area of concern
would be whether significant errors in recovering $f(r)$ appear
unless the estimate function for $h(r)$ is highly accurate; one can
investigate whether this is so by running the extraction of $f(r)$
several times using a different estimate for $h(r)$ each time. We
see how this affects the results in the first example below.
Finally, one should note that the depth to which the metric can be
recovered is subject to the same restrictions as before (see section
\ref{sec:singspaces}): for example in singular spacetimes, whilst
the spacelike geodesics can probe down to the horizon radius,
$r_{h}$ (and we thus obtain $h(r)$ down to $h(r_{h})$), the null
geodesics can only probe as far as the first local maximum in the
effective potential (figure \ref{Mv1}), at some $r_{h2} > r_{h}$,
leaving $f(r)$ undetermined for $r < r_{h2}$. Nevertheless, by
combining the two different approaches to probing the bulk, we have
obtained more information than is possible using either
individually.

\subsection{Example 1: Testing the combined extraction procedure}

Consider a spacetime where the two metric functions $f(r)$ and
$h(r)$ are given by the following:

\begin{equation} \label{eq:fexnew1}
f(r) = 1 + r^{2} - \frac{7 \, r^{2}}{(r^{2} + 1)(r^{2} + 13)} +
\frac{2 r \sin(5 \, r)}{r^{4} + 15}
\end{equation}
\begin{equation} \label{eq:hexnew1}
h(r) = \left(1 + r^{2} - \frac{4 \, r^{2}}{(r^{2} + 1)(r^{2} + 8)} +
\frac{3 r \sin(2 \, r)}{r^{4} + 1}\right)^{-1}
\end{equation}

Whilst this is in no way meant to be a representation of any
physical deformation of the bulk, it is a good test of the combined
extraction method, as it provides a monotonic effective potential
for the null geodesics, and so allows us to probe down to $r = 0$.
One can also use the similarity between this spacetime and that
described in the first example of section \ref{sec:examples}, namely
that we have $h(r) = f_{1}(r)^{-1}$. This was deliberately chosen so
the part of the metric probed by the spacelike geodesics is exactly
as it was in the case of example 1; the change in $f(r)$ has no
effect on the results, and thus the best fit estimates for $h(r)$
are exactly those specified by the values of the parameters in Table
\ref{table11}. We therefore have four different estimates for $h(r)$
(one for each of the four choices of step size used), and we label
them $h_{0.1}(r)$ through to $h_{0.005}(r)$, where the subscript
refers to the step size. All that is left to do is to attempt to
recover $f(r)$ via the null geodesic data\footnote{As we saw in
\cite{hammer}, one can use a range of different step sizes in $y$ to
obtain varied levels of accuracy in the metric extraction; as we are
not intending to specifically analyze the null geodesic method here,
we simply choose a starting value of $y = 0.9985$, and a step size
of $\triangle y = 0.0005$, as these are sensible values for the
example given. \label{foot1}} for each fit to $h(r)$, and compare it
firstly to the actual values, and also to those obtained using the
exact function $h(r)$ rather than an estimate. The results are
analyzed using a best fit of the form of \eqref{eq:ffit11} and are
presented in Table \ref{table44}.

\begin{table}
\begin{center}
\begin{tabular}{l l l l l l l}
\hline $h_{\textrm{fit}}(r)$ & $\alpha$ (7)& $\beta$ (1)& $\gamma$ (13)& $\chi$ (2)& $\eta$ (5)& $\lambda$ (15)\\
\hline $h_{0.1}(r)$ & 6.81 & 1.03 & 12.49 & 2.00 & 4.99 & 14.92 \\
$h_{0.05}(r)$ & 6.81 & 1.03 & 12.48 & 2.00 & 4.99 & 14.92 \\
$h_{0.01}(r)$ & 6.80 & 1.03 & 12.48 & 2.00 & 4.99 & 14.92 \\
$h_{0.005}(r)$ & 6.80 & 1.03 & 12.48 & 2.00 & 4.99 & 14.92 \\
$h(r)$ & 6.80 & 1.03 & 12.48 & 2.00 & 4.99 & 14.92 \\
\hline
\end{tabular}
\end{center}
\caption{Best fit values (to 2 d.p.) for the $f_{\textrm{fit}}(r)$
parameters $\alpha$, $\beta$, $\gamma$, $\chi$, $\eta$ and
$\lambda$, with the actual values indicated in brackets. We see that
even our roughest estimate for $h(r)$ is close enough for the
extraction of $f(r)$ to be highly accurate.}\label{table44}
\end{table}
We see quite clearly from the table of results that even using our
roughest estimate for $h(r)$, namely $h_{0.1}(r)$, we obtain a
highly accurate estimate for $f(r)$. Indeed, the limiting factor is
not the accuracy of the estimate for $h(r)$, rather it is the choice
of step size and starting $y$ in the null geodesic part of the
extraction (see footnote \ref{foot1}).

\subsection{Example 2: Radiation in $AdS_{3}$, a toy model}

\begin{figure}
\begin{center}
  \includegraphics[width=0.9\textwidth]{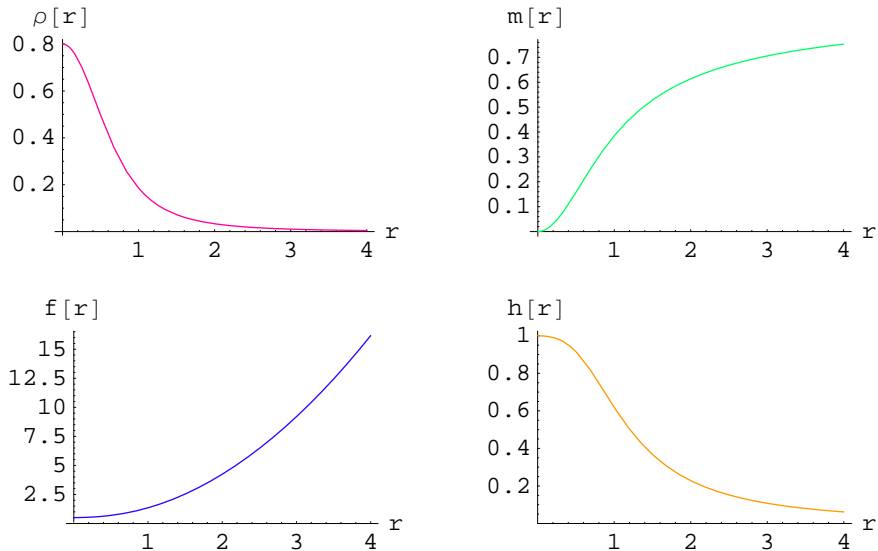}\\
\end{center}
\caption{The density and mass profiles (top plots) for a ``star''
with central density $\rho_{0} = 0.8$, along with plots of the
corresponding metric functions $f(r)$ and $h(r)$
(bottom).}\label{star1}
\end{figure}

As the two extraction methods give such good fits when applied
sequentially, we now turn our attention to a less trivial example,
where we consider a gas of radiation in $AdS_{3}$. There have been
numerous papers exploring this and other closely related geometries
in various dimensions, such as \cite{vero,page,radu1,radu2}, and we
focus here purely on our ability to recover the metric information
via our numerical extraction methods. Firstly, we note that whilst
restricting ourselves to three bulk dimensions does make our
spacelike geodesic method fully applicable (see section
\ref{sec:dims}), it also restricts the physical realism of the model
due to the non-dynamical nature of gravity. Nevertheless, it
provides a good toy model for radiating ``stars'' in AdS spacetimes,
and allows us to demonstrate how well the pertinent information
(e.g. the ``star's'' mass and density profiles) about the bulk can
be recovered. We consider a perfect fluid solution to Einstein's
equations, with the pressure $P(r)$ set equal to half the density,
$\rho(r)/2$, as for radiating matter the stress-energy tensor is
traceless. For a metric of the form of \eqref{eq:newAdSmetric}, we
find that\footnote{We set $R = 1$ and $8 \pi G_{3} \equiv 1$ for
convenience.}:

\begin{equation} \label{eq:starhr}
h(r) = \left(1 + r^{2} - m(r)\right)^{-1}
\end{equation}
and
\begin{equation} \label{eq:starfr}
f(r) = \left(\frac{\rho_{\infty}}{\rho(r)} \right)^{2/3}
\end{equation}
where the mass function is defined by:
\begin{equation} \label{eq:starmr}
m(r) = 2 \int_{0}^{r} \rho(\acute{r}) \acute{r} \, \mathrm{d}
\acute{r}
\end{equation}
and $\rho_{\infty}$ is the leading coefficient of $\rho(r)$ at large
$r$, and is given by $\rho_{\infty} \approx \rho(r) r^{3}$ as $r
\rightarrow \infty$. We obtain from the field equations a pair of
coupled ODEs for $m(r)$ and $\rho(r)$:

\begin{equation} \label{eq:starode1}
m'(r) = 2 \rho(r) r
\end{equation}
\begin{equation} \label{eq:starode2}
\frac{6 + 3 \rho(r)}{1 + r^{2} - m(r)} + \frac{2 \rho'(r)}{\rho(r)
r} = 0
\end{equation}
which when combined with the relevant boundary conditions $m(0) = 0$
and $\rho(0) = \rho_{0}$ can be numerically solved to allow us to
generate the geometry of the spacetime (see figure \ref{star1}). The
condition $\rho(0) = \rho_{0}$ specifies the internal density of the
gas, and $\rho_{0}$ is the single free parameter of the system: pure
AdS is recovered when $\rho_{0} = 0$.

\begin{figure}
\begin{center}
  \includegraphics[width=0.9\textwidth]{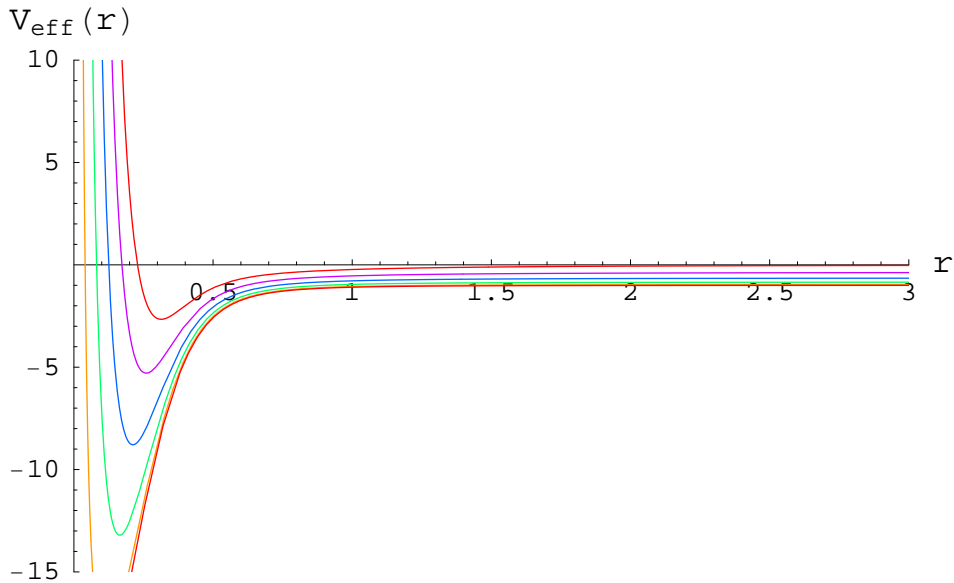}\\
\end{center}
\caption{Effective potentials for null geodesics in a spacetime with
$M = 8$. The upper (red) potential is for $y \equiv J/E = 0.9999$;
no matter how close to one the ratio $J/E$ becomes, the minimum
radius (defined by $V_{eff} = 0$) remains
small.}\label{starpotentials}
\end{figure}

Before we begin with the metric extraction, we should make a comment
about the features of such spacetime at large radius, as there are
significant differences in the asymptotic behaviour of the metric
depending on the choice of $\rho_{0}$. For $\rho_{0} \ne 0$, we have
that the asymptotic behaviour of the metric functions is given by

\begin{equation} \label{eq:starasymp}
h(r) \rightarrow \left(1 + r^{2} - M\right)^{-1} \;\;\;\;
\textrm{and} \;\;\;\; f(r) \rightarrow 1 + r^{2} - M \;\;\;\;\;\;
\textrm{as } \, r \rightarrow \infty
\end{equation}
where $M > 0$ is a constant. If $M > 1$ we have that the metric
becomes the BTZ black hole solution at large $r$ (see \cite[for
example]{banks,mats} for more details); this poses a problem for the
method involving null geodesics, as we can no longer use them to
probe the full range of $r$. Whilst this is due to the form of the
effective potential (see figure \ref{starpotentials}), it is not due
to the local maximum problem we saw in section \ref{sec:singspaces}.
Rather here we no longer have geodesics which can usefully probe the
spacetime \textit{away} from the centre: for the full set of null
geodesics (obtained by varying $y$ for zero to one), the minimum
radius reached by the geodesics is bounded from above. We thus
cannot take $r_{min}$ to be arbitrarily large on the first step of
our iteration, which was necessary for us to begin extracting the
metric (although we should note that we could still apply the
spacelike method to extract $h(r)$ in this scenario). Instead
however, we will consider the region $0 < M < 1$, corresponding to
conical defects, in which both methods are applicable and is
obtained by taking $\rho_{0}$ to be small.\footnote{One should also
note from \eqref{eq:starasymp} that our iterative equations for
recovering the metric need to be modified to take into account the
new asymptotic behaviour, as we no longer have that the metric is
given by $f(r) \approx r^{2} + 1$ at large $r$. Thus we say that for
$r \ge r_{n}$ we have that $f(r)$ and $h(r)$ are given by
\eqref{eq:starasymp}, and modify the approximations to the integrals
for $\phi_{n-i}$ and $\mathcal{L}_{n-i}$ accordingly.}

\begin{figure}
\begin{center}
  \includegraphics[width=0.48\textwidth]{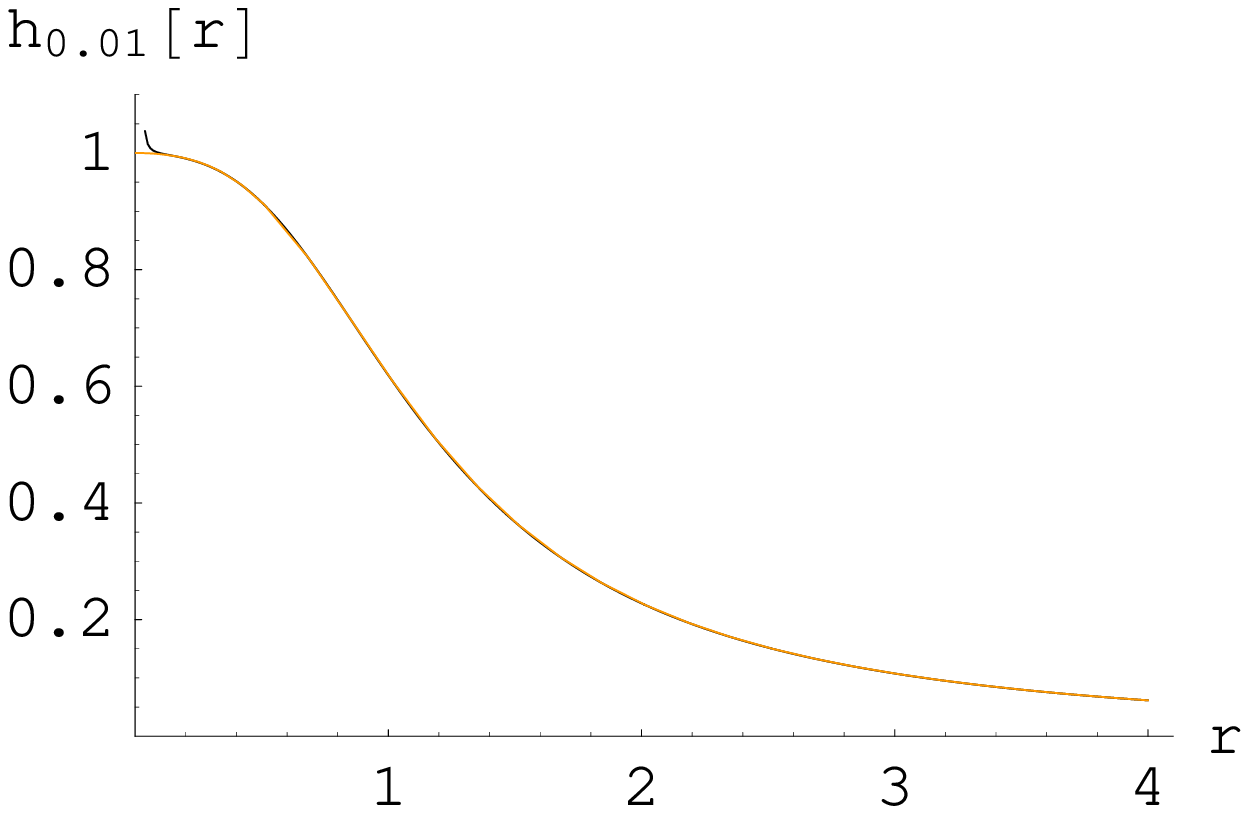}
  \includegraphics[width=0.48\textwidth]{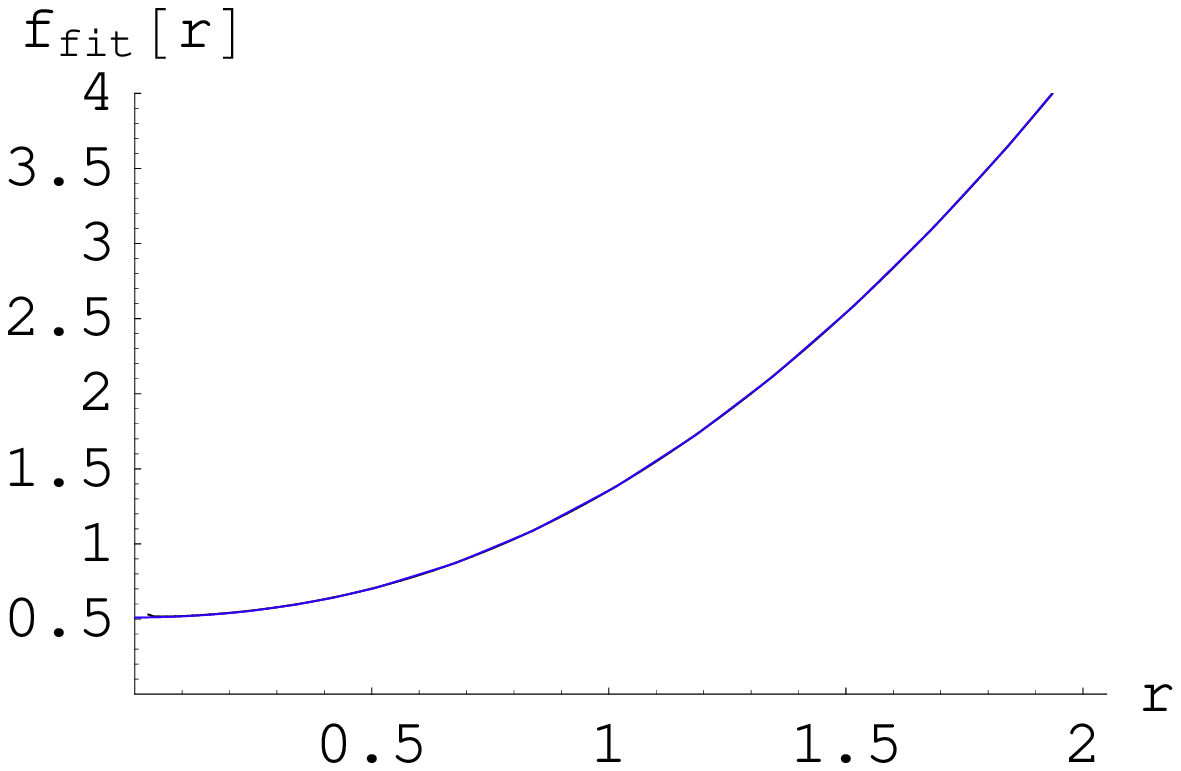}\\
\end{center}
\caption{The third (and most accurate) estimate for $h(r)$, where
the fit is good down to $r \sim 0.1$ (left plot). The estimate for
$f(r)$ generating using this approximation to $h(r)$ is given in the
right plot, and we see that it too appears accurate down to very low
$r$.}\label{star3}
\end{figure}

Let us then proceed with recovering the metric in the specific
example shown in figure \ref{star1}, where we have set $\rho_{0} =
0.8$. Bearing in mind that our goal is to firstly reconstruct the
functions $f(r)$ and $h(r)$, and then use these to determine the
mass and density profiles ($m(r)$ and $\rho(r)$ respectively) of the
star, we begin by applying the spacelike geodesic method (with step
sizes of $0.1$, $0.05$ and $0.01$) to produce three estimates for
$h(r)$, the most accurate of which, namely $h_\textrm{0.01}(r)$, is
shown in figure \ref{star3}. Whilst in the previous example we
defined $h(r)$ explicitly by hand, and so knew the form of the
function with which to apply the non-linear fit to generate the best
fit curve $h_\textrm{fit}(r)$, here we do not have such a starting
point. Instead, we use the data points $(r_{n-i},h(r_{n-i}))$ to
generate an interpolating function which will serve as our
$h_{\textrm{fit}}(r)$. Thus although we cannot write down an
explicit form for $h_{\textrm{fit}}(r)$, we can use the
interpolating function to then carry out the next part of the
extraction process, namely using the null geodesic probes to recover
$f(r)$.

Using the third (and most accurate) estimate for $h(r)$ in the
modified null geodesic method of section \ref{sec:combo} and
Appendix C, we produce the estimate for $f(r)$,
$f_{\textrm{fit}}(r)$, also shown in figure \ref{star3}: we have now
reconstructed the star metric. Although if we so wished we could
have taken smaller step sizes to improve both the estimate of $h(r)$
and that of $f(r)$, we now continue with the ones we have.

\begin{figure}
\begin{center}
  \includegraphics[width=0.9\textwidth]{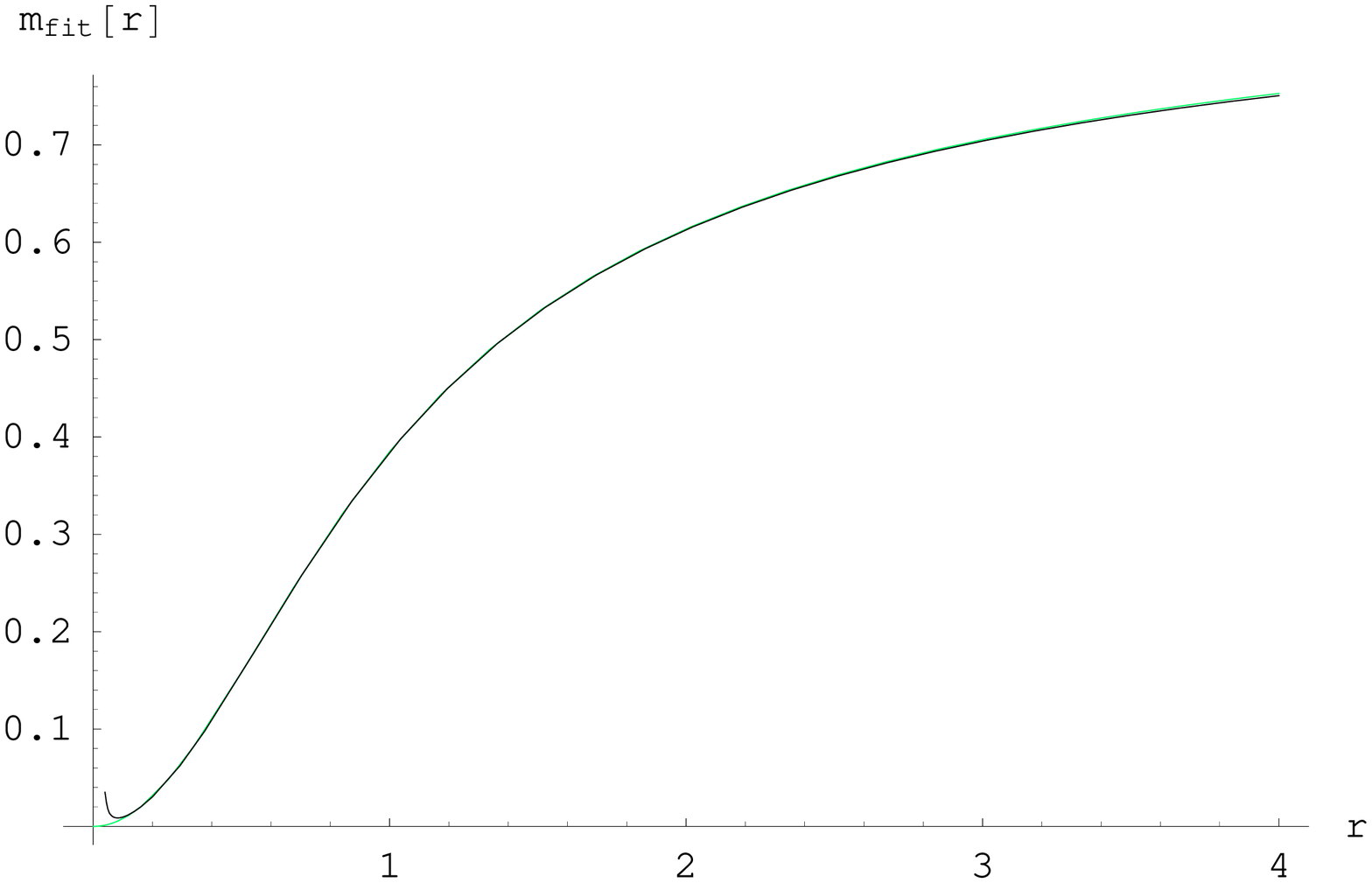} \\
  \includegraphics[width=0.9\textwidth]{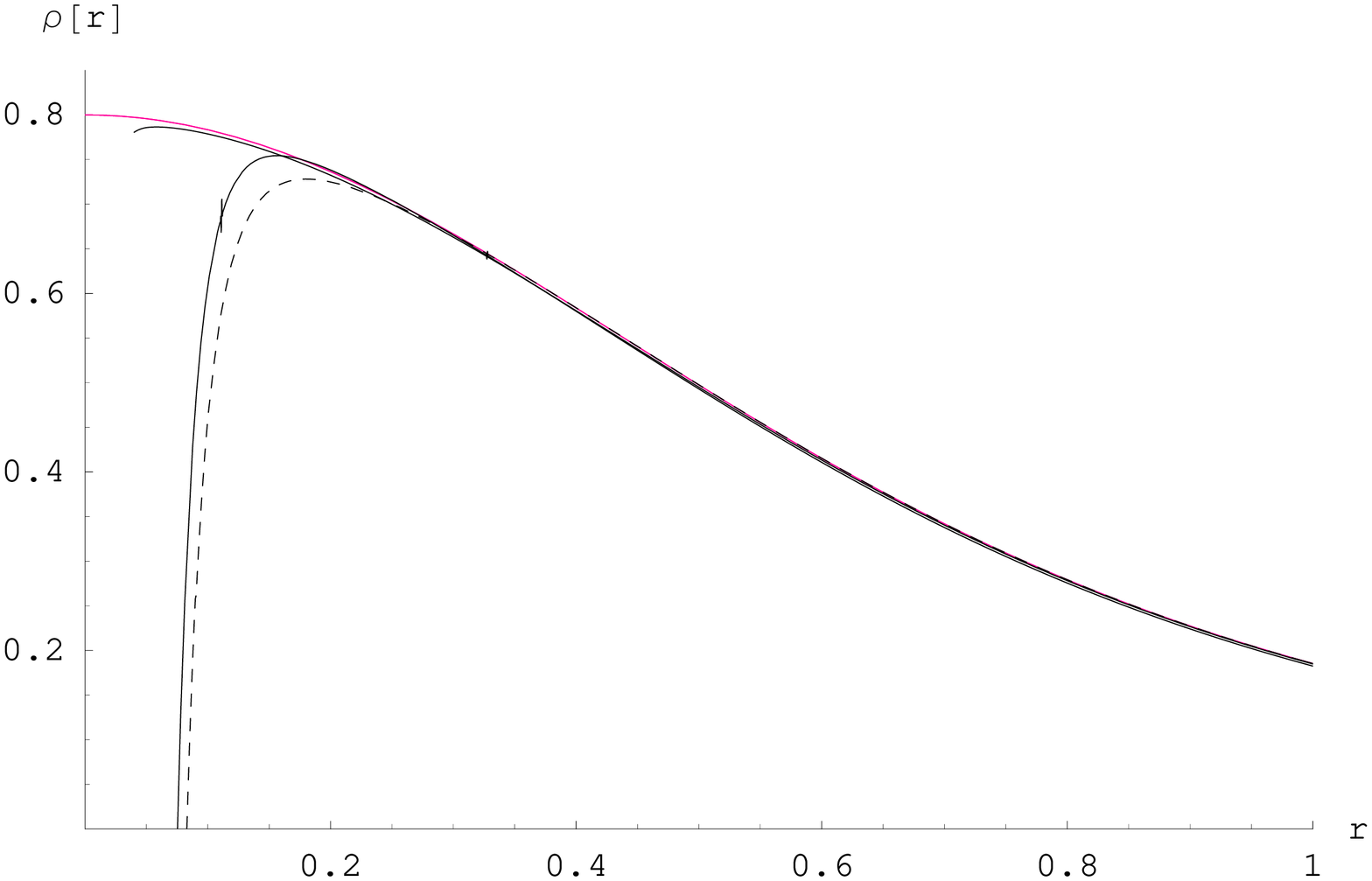}\\
\end{center}
\caption{Estimates for the mass and density profiles for our
``star''. As with $h_{\textrm{0.01}}(r)$ and $f_{\textrm{fit}}(r)$,
these match the actual curves closely until low $r$, although the
density estimate $\rho_{\textrm{fit}}(r)$ (dashed) fails at
noticeably higher $r$ than the others. Included in the lower plot
are alternative estimates for for the density profile, obtained from
\eqref{eq:starfr} (closest fit) and \eqref{eq:starpfit}
(solid)}\label{star5}
\end{figure}

How do we use the metric functions to determine the mass and density
information for the star? From \eqref{eq:starhr} it is immediately
obvious: we can rearrange the equation to solve for $m(r)$, and
substitute in our interpolating function $h_{\textrm{fit}}(r)$ to
give an estimate for the mass profile:
\begin{equation} \label{eq:starmfit}
m_{\textrm{fit}}(r) = 1 + r^{2} - \frac{1}{h_{\textrm{fit}}(r)}
\end{equation}
and we obtain a fit for the density profile in similar fashion, by
using the above estimate for $m(r)$ in \eqref{eq:starode1}, to give:
\begin{equation} \label{eq:starrhofit}
\rho_{\textrm{fit}}(r) = \frac{m_{\textrm{fit}}'(r)}{2 r}
\end{equation}
These two fits are plotted against the actual functions $m(r)$ and
$\rho(r)$ in figure \ref{star5}, and we see that by using the metric
function data $h_{\textrm{fit}}(r)$ we have obtained reasonably good
estimates of the mass and density profiles of the star, aside from
at very small $r$, where the errors from the estimate of $h(r)$
become noticable. What is noticeable is that the estimate for
$\rho(r)$ fails at higher $r$ than any of the others; this is due to
the use of the derivative of the interpolating function
$m_{\textrm{fit}}(r)$ in its construction, and is dealt with later
(see below).

One now asks the obvious question of why it was necessary to extract
the function $f(r)$ at all, seeing as we have apparently just
reconstructed the information about the star simply by using
$h_{\textrm{fit}}(r)$. This is where we recall that we should be
assuming that \textit{a priori} we knew nothing about the origin of
the metric's deviation from pure AdS. In fact, this has not been the
case. Whilst our expressions for $h(r)$ in terms of $m(r)$ and the
mass $m(r)$ in terms of the density $\rho(r)$, \eqref{eq:starhr} and
\eqref{eq:starode1}, stem from the dimensionality of the bulk (e.g.
in higher dimensions one would have the $m(r)$ term multiplied by
some negative power of $r$), in defining $f(r)$ by \eqref{eq:starfr}
we have already taken the matter content to be a gas of radiation,
which sets $P(r) = \rho(r)/2$ and removes the pressure profile as an
unknown. Given this knowledge, one could indeed have simply used the
spacelike geodesic method say to extract the information about the
star, as $h(r)$ gives $m(r)$, and $m(r)$ gives $\rho(r)$. Extracting
$f(r)$ becomes a necessity, however, if one removes the assumption
about the matter content; then one also has to compute the pressure
profile. It is most easily determined (once we have our fits for
$f(r)$ and $h(r)$) from the $G_{r r}$ component of Einstein's
equations, and we have that:
\begin{equation} \label{eq:starpfit}
P_{\textrm{fit}}(r) = \frac{f_{\textrm{fit}}'(r)}{2 \, r
f_{\textrm{fit}}(r) h_{\textrm{fit}}(r)} - 1
\end{equation}
which in our example corresponds to $\rho_{\textrm{fit}}(r)/2$.
Therefore by also plotting $2 \, P_{\textrm{fit}}(r)$ in figure
\ref{star5}, we can see how close the fits generated by the two
different expressions \eqref{eq:starrhofit} and \eqref{eq:starpfit}
match, and this provides a check that the matter content is indeed
that of a gas of radiation and confirms that our expression,
\eqref{eq:starfr}, for $f(r)$ is correct. Interestingly, we see that
this expression provides a slightly better fit to $\rho(r)$ at small
$r$ than that from \eqref{eq:starrhofit}. This is simply because
\eqref{eq:starpfit} includes $f_{\textrm{fit}}(r)$ terms, and the
non-linear step size in $r$ in the null extraction method generates
a greater amount of data points at low $r$ for the estimate for
$f(r)$, thus allowing the derivative of the interpolation function
to be more accurately determined. We can obtain the best fit at low
$r$ by using $f_{\textrm{fit}}(r)$ in \eqref{eq:starfr} and solving
for $\rho(r)$ (see figure \ref{star5}), where we have avoided using
derivatives.\footnote{One should note this does firstly require the
value of $\rho_{\infty}$ to be determined from the fall off of
$\rho(r)$ at large $r$; this is however available from our earlier
fit to $\rho$ given in \eqref{eq:starrhofit}.}

Finally, we can use the estimates to give a numerical value for our
free parameter $\rho_{0}$. Taking $\rho_{\infty}$ as having been
calculated from the asymptotic fall off, and approximating the value
of $f(0)$ as $0.525$,  we obtain a value of $0.76$, compared with
the actual value of $\rho_{0} = 0.8$. Whilst the match is fairly
good, this is where the accuracy of the estimates for $f(r)$ and
$h(r)$ become very important; in taking $f(0) = 0.525$ we have
discarded the final few iterations of $f_{\textrm{fit}}(r)$ at small
$r$, which lead to a kink in the curve, as being erroneous and due
to an incomplete recovery of $h(r)$. This is a reasonable assumption
to make, as in our previous examples we saw that for too large a
step size the method of generating $h_{\textrm{fit}}(r)$ fails to
reach down to $r = 0$. We also have the data from the higher step
size fits ($h_{\textrm{0.1}}(r)$ and $h_{\textrm{0.05}}(r)$) with
which to analyse the accuracy of our estimates for $h(r)$ at low
$r$. However, as it is the small $r$ region from which the numerical
value of $\rho_{0}$ is calculated, in order for it to be confidently
extracted one must ensure the estimates $h_{\textrm{fit}}(r)$ and
$f_{\textrm{fit}}(r)$ are thoroughly checked for $r$ close to zero.

\section{Extensions to less symmetric cases} \label{sec:extension}

In all of the above we have taken the spacetime metric to be both
static and spherically symmetric, however, we now consider how the
methods for extracting the bulk presented here could be extended to
include more general cases.

Reducing the amount of symmetry removes conserved quantities from
the geodesics; spherical symmetry gives us conservation of angular
momentum, time translational symmetry gives us energy conservation.
Consequently, there will be additional unknowns introduced in our
analysis of the geodesic path, as we will need to know more details
about its route through the bulk; this should not prove a problem,
however, as there will also be further information available from
the geodesic equations.

Consider for example the non-spherically symmetric (but still
static) case. Before, when there was no angular dependence in the
metric, we considered a series of geodesics which probed deeper and
deeper into the bulk - in other words, we had one which probed down
to each $r_{n-i}$. These were specified by the angular separation of
the endpoints on the boundary, and the actual values of the
$\phi_{start}$ and $\phi_{end}$ were unimportant. This allowed us to
reconstruct the bulk step by step, one value of $f(r_{n-i})$ at a
time.

Now, what is the analogous method in the non-spherically symmetric
case? At each step of the iteration we can still consider some fixed
angular separation of the endpoints, however, we must also vary
$\phi_{start}$ from $0$ to $2 \pi$ (with some choice of slicing
sufficient to give an accurate estimate), such that for each
iterative step we recover a ``ring'' of information about the
metric. The subsequent steps then recover smaller and smaller rings,
extracting the metric function down to the centre of the spacetime.
This is the basic idea of the extraction method; finalising a more
detailed procedure which gives high accuracy within a reasonable
computational time is subject of current research, one now has two
step sizes to consider: the slicing in $\phi$ and the radial steps
in $r$.

Finally, we should recall that higher dimensional cases offer
further complications, as mentioned in section \ref{sec:dims}, as
although the null geodesic method is already applicable in such
cases, the spacelike method is not. In principle though, the ideas
still hold; one would need an expression for the correct minimal
surface corresponding to the entanglement entropy (see \cite{hubtak}
for more details on this point) which could then be treated in much
the same way as the geodesic proper length, as they will each probe
to a certain depth in the bulk, and those remaining at large $r$
will behave as in pure AdS. Completing the analysis for these cases
is again the subject of further research.

\section{Discussion} \label{sec:conc}

In this paper we have seen how the bulk geometry can be extracted
(in asymptotically Anti-de Sitter spacetimes) using the entanglement
entropy information obtained from the corresponding boundary field
theory. In the three dimensional case, the entanglement entropy of a
subsection $A$ of the $1+1$ dimensional boundary is proportional to
the proper length of the static spacelike geodesic connecting the
endpoints of $A$ (see figure \ref{stat1}). Using this relation,
\eqref{eq:entrop1}, together with the endpoint data allows both the
minimum radius, $r_{min}$, of the spacelike geodesic and the value
of the metric function $f(r_{min})$ at this point to be determined,
provided sufficient information about the bulk is known for $r >
r_{min}$. Thus by starting from large $r$, where the metric is
approximately pure AdS, one can probe further and further into the
bulk using geodesics connecting progressively smaller regions on the
boundary.

This extraction of the metric is made significantly more
straightforward by an interesting relationship between the proper
length of the geodesic and the angular separation of its endpoints.
Specifically, the gradient, $d \mathcal{L}/d \phi$, gives the
angular momentum, $J$, of the corresponding geodesic, which in the
static, spherically symmetric cases considered here, is equal to the
minimum radius the geodesic probes down to in the bulk.

An iterative method for recovering the metric information in
practice by approximating the relevant geodesic equation was thus
then proposed, and a number of examples were given. The iterative
method was developed in analogous way to the method presented in
\cite{hammer}, which used the endpoint data of null geodesics to
similarly extract the bulk information, and was reviewed in section
\ref{sec:review}.

A comparison of the two methods was then made, investigating their
relative advantages and disadvantages; this highlighted a number of
differences in their relative applicabilities. Whilst the method of
\cite{hammer}, which uses null geodesics as probes, runs into
problems when encountering geometries with significant deviation
from pure AdS (as this leads to a non-monotonic effective potential
for the geodesics which limits the depths to which the metric
information can be recovered), no such limitations arise for the
method given here involving spacelike geodesics, unless the metric
is also singular. On the other hand, the null geodesic method is
completely applicable in any number of dimensions, whereas although
the principle of extracting the metric via spacelike geodesics can
be extended to greater than three bulk dimensions, the relation
between entanglement entropy and minimal surface area
\eqref{eq:entrop1} no longer involves the geodesic's proper length,
and thus this quantity is no longer so readily available from the
CFT. Computationally, the two methods (as presented here) are of
comparable efficiency, although both have scope for further
optimization.

Significantly, we demonstrated in section \ref{sec:combo} how the
two methods can be applied together to allow the probing of the most
general static, spherically symmetric asymptotically AdS spacetimes,
with metric of the form of \eqref{eq:newAdSmetric}. This is a
notable extension to the applicability of either method
individually, as whilst part of the metric information (i.e. the
$h(r)$ of \eqref{eq:newAdSmetric}) could always be extracted using
the spacelike geodesics, they could never give any insight into
$f(r)$. The null geodesics, on the other hand, can probe both $f(r)$
and $h(r)$ but without yielding enough information to determine
either, without some \textit{a priori} knowledge of a relationship
between them. It is the separation of the two functions in the
spacelike case, however, which allows the methods to be combined so
straightforwardly, as one firstly determines an estimate for $h(r)$,
then takes this as a known function when analyzing the null geodesic
data. We concluded by considering the toy model scenario of a gas of
radiation (a ``star'') in $AdS_{3}$ and demonstrated how the
recovery of the metric allowed the pertinent information of the star
(its mass and density profiles) to be well estimated. Whilst the
estimates produced were accurate down to low $r$ (dependent on the
choice of step size in both the null and spacelike methods), one had
to be careful when using the derivative of $m_{\textrm{fit}}(r)$
(the interpolating fit to the mass profile) to generate
$\rho_{\textrm{fit}}(r)$. Although the fit produced was still good,
it failed at noticeably larger $r$ than the fits for any of the
other functions, due to inaccuracies introduced via
$m_{\textrm{fit}}'(r)$. This could be avoided by using the
alternative definition of $\rho(r)$ in terms of $f(r)$,
\eqref{eq:starfr}, provided one first extracted the asymptotic fall
off of the density as $r \rightarrow \infty$.

Finally, we noted in section \ref{sec:extension} that this work can
be extended further by considering spacetimes which are not
spherically symmetric, and by investigating the higher dimensional
cases where the area of the minimal surface relating to the
entanglement entropy is not simply the length of the corresponding
static spacelike geodesic. Both avenues have the prospect of
yielding fruitful results for metric computation in AdS/CFT. Also,
one could approach the problem of metric extraction from a different
angle, by investigating whether there are alternative methods
available which do not involve the use of geodesic probes. If so, it
would be interesting to see whether these lead to more efficient
ways of computing the metric functions than those described here.

\section*{Acknowledgements}

For useful discussions and feedback I wish to thank Veronika Hubeny
(who also provided much encouragement and helpful information) along
with Simon Creek, Martyn Gigg, Elizabeth Holman, Kemal Ozeren and
James Umpleby. This work was supported by an EPSRC studentship grant
and the University of Durham Department of Mathematical Sciences.

\section*{Appendix A}

In section \ref{sec:method1} we outlined the principle behind the
iterative technique of metric extraction: determining $r_{min}$ from
the gradient $d \mathcal{L} / d \phi$ and then calculating an
estimate for $f(r_{min})$ at each step by splitting up the integral
in \eqref{eq:lphi4space} and approximating each piece separately,
beginning the whole process at large $r$, where the metric is
approximately pure AdS and we can take $f(r) \approx r^{2} + 1$.
Here we go on to give further details of how to set this up, and
explicitly write down the equations used in the
approximations\footnote{The procedure used here is only one of a
number of possible methods for discretizing the integral; for the
purposes of illustrating the principle, this method is both brief
and accurate to a good degree.}.

Having taken the first step which chooses an $r_{n}$ large enough so
the metric is approximately pure AdS, and hence $f(r_{n}) =
r_{n}^{2} + 1$, we can continue as follows. For a geodesic with
slightly lower angular momentum $J_{n-1}$ (which can be obtained by
taking a slightly larger region B on the boundary), we can split up
the integral over $r$ in \eqref{eq:phir} into two pieces:
\begin{equation} \label{eq:phir3}
\phi_{n-1} = 2 \int_{r_{n-1}}^{r_{n}} \frac{1}{r \sqrt{f(r)}
\sqrt{\frac{r^{2}}{J^{2}} - 1}} \, \mathrm{d} r + 2
\int_{r_{n}}^{r_{max}} \frac{1}{r \sqrt{f(r)}
\sqrt{\frac{r^{2}}{J^{2}} - 1}} \, \mathrm{d} r
\end{equation}

The first integral in the above can be well approximated by taking a
next-to-lowest order series expansion about the point $r = r_{n-1}
\left( = J_{n-1} \right)$, as the distance $r_{n} - r_{n-1}$ is
small. For the second integral, we can again use our assumption that
$f(r) = r^{2} + 1$ for $r \ge r_{n}$, and overall we obtain for the
angular separation of the endpoints:

\begin{eqnarray} \label{eq:phir4}
\phi_{n-1} \approx && 2 \sqrt{2} \sqrt{\frac{r_{n} -
r_{n-1}}{r_{n-1} \, f(r_{n-1})}} - \frac{5 f(r_{n-1}) + 2 \, r_{n-1}
f'(r_{n-1})}{3 \sqrt{2}} \left(\frac{r_{n} - r_{n-1}}{r_{n-1} \,
f(r_{n-1})}\right)^{3/2} \nonumber \\ &&  + \arctan{\left(\frac{2
r_{n-1}^{2} + \left(r_{n-1}^{2} - 1\right)r_{n}^{2}}{2 r_{n-1}
\sqrt{r_{n}^{4} - \left(r_{n-1}^{2} - 1\right)r_{n}^{2} -
r_{n-1}^{2}}}\right)} \nonumber \\ && -
\arctan{\left(\frac{r_{n-1}^{2} - 1}{2 r_{n-1}}\right)}
\end{eqnarray}
where we have again taken the limit $r_{max} \gg r_{n} > r_{n-1}$.
Alternatively, one could perform similar approximations on the
equation for the proper length, \eqref{eq:lengthr}, to obtain:

\begin{eqnarray} \label{eq:lengthr4}
\mathcal{L}_{n-1} \approx && 2 \sqrt{2 r_{n-1}} \sqrt{\frac{r_{n} -
r_{n-1}}{f(r_{n-1})}} + \frac{3 f(r_{n-1}) - 2 \, r_{n-1}
f'(r_{n-1})}{3 \sqrt{2 \, r_{n-1}}} \left(\frac{r_{n} - r_{n-1}}{
f(r_{n-1})}\right)^{3/2} \nonumber \\ && + 2 \log{\left(2 \,
r_{max}\right)} - 2 \log{\left( \sqrt{r_{n}^{2}-r_{n-1}^{2}} +
\sqrt{r_{n}^{2}+1}\right)}
\end{eqnarray}
In the above expressions we have introduced a further unknown,
namely the gradient of the function $f(r)$ at the point $r =
r_{n-1}$; this can be eliminated by taking the simple linear
approximation:

\begin{equation} \label{eq:fgradm1}
f'(r_{n-1}) \approx \frac{f(r_{n}) - f(r_{n-1})}{r_{n} - r_{n-1}}
\end{equation}
which holds provided the radial distance $r_{n} - r_{n-1}$ is kept
small.\footnote{The presence of an $f'(r)$ term deserves further
comment: one can avoid introducing it by using the lowest order
expansion, however, this reduces the overall accuracy of the method.
The detrimental effect of the approximation to the gradient on the
accuracy of the estimates is not as pronounced as in the method of
\cite{hammer} due to the use of linear step size in $J$ (and hence
$r$), see section \ref{sec:examples}. Surprisingly, an alternative
integral one might consider when setting up the iteration, which
allows the higher order series expansion to be used without
introducing $f'(r)$, leads to an unstable method rather than a more
accurate one, see Appendix B.} As we mentioned in section
\ref{sec:entangle}, we can calculate the proper length
$\mathcal{L}_{n-1}$ from the relevant entanglement entropy
expression, in general this is given by:

\begin{equation} \label{eq:entrop2}
\mathcal{L} = 4 \, G_{N}^{(3)} S_{A}
\end{equation}
where $S_{A}$ corresponds to the entanglement entropy for the
subsystem A in the deformed spacetime. Taking the entanglement
entropy as a known quantity from the CFT, along with the angular
separation of the endpoints (which is given simply from the length
of the subsystem in the CFT) one can calculate the corresponding
minimum radius $r_{n-1}$ from \eqref{eq:dldphibspace}, and so our
only remaining unknown in both \eqref{eq:phir4} and
\eqref{eq:lengthr4} is $f(r_{n-1})$. We can thus numerically solve
either for $f(r_{n-1})$, and determine the metric function at this
point. Continuing in a similar fashion, by taking geodesics with
progressively smaller angular momenta and numerically solving at
each step, we can iteratively extract the complete metric. For
general $\phi_{n-i}$ and $\mathcal{L}_{n-i}$ the integrals are split
up into $(i+1)$ pieces; two are approximated as in \eqref{eq:phir4}
and \eqref{eq:lengthr4}, with the remaining terms evaluated using
Simpson's rule (a polynomial fit to the curve). The general
expression for $\phi_{n-i}$ can then be written as:

\begin{equation} \label{eq:phir5}
\phi_{n-i} \approx A_{n-i} + B_{n-i} + C_{n-i}
\end{equation}
where
\begin{equation} \label{eq:anminusi}
A_{n-i} = 2 \sqrt{2} \sqrt{\frac{r_{n-i+1} - r_{n-i}}{r_{n-i} \,
f(r_{n-i})}} - \frac{5 f(r_{n-i}) + 2 \, r_{n-i} f'(r_{n-i})}{3
\sqrt{2}} \left(\frac{r_{n-i+1} - r_{n-i}}{r_{n-i} \,
f(r_{n-i})}\right)^{3/2}
\end{equation}
\begin{equation} \label{eq:cnminusi}
C_{n-i} = \arctan{\left(\frac{2 r_{n-i}^{2} + \left(r_{n-i}^{2} -
1\right)r_{n}^{2}}{2 r_{n-i} \sqrt{r_{n}^{4} - \left(r_{n-i}^{2} -
1\right)r_{n}^{2} - r_{n-i}^{2}}}\right)} -
\arctan{\left(\frac{r_{n-i}^{2} - 1}{2 r_{n-i}}\right)}
\end{equation}
are the two approximations we had before, and the $B_{n-i}$ term is
given by:
\begin{equation} \label{eq:bnminusieven}
B_{n-i} = \sum_{j=1}^{i/2} \frac{\left(r_{n-2j+3} -
r_{n-2j+1}\right)}{3} \left(g_{n-i}(r_{n-2j+3})+ 4 \,
g_{n-i}(r_{n-2j+2}) +g_{n-i}(r_{n-2j+1}) \right)
\end{equation}
for $i$ even\footnote{Using this definition requires a value for the
$r_{n+1}$ term, which can be obtained in an identical way to that
used in determining $r_{n}$}, and by
\begin{equation} \label{eq:bnminusiodd}
B_{n-i} = \sum_{j=1}^{(i-1)/2} \frac{\left(r_{n-2j+2} -
r_{n-2j}\right)}{3} \left(g_{n-i}(r_{n-2j+2})+ 4 \,
g_{n-i}(r_{n-2j+1}) +g_{n-i}(r_{n-2j}) \right)
\end{equation}
for $i$ odd, where we have defined the function
\begin{equation} \label{eq:newgdef1}
g_{n-i}(r) \equiv \frac{1}{r
\sqrt{f(r)}\sqrt{\frac{r^{2}}{r_{n-i}^{2}}-1}}
\end{equation}
for ease of notation. For the proper length we similarly have that:

\begin{equation} \label{eq:lengthr5}
\mathcal{L}_{n-i} \approx \mathcal{A}_{n-i} + \mathcal{B}_{n-i} +
\mathcal{C}_{n-i}
\end{equation}
with
\begin{equation} \label{eq:aanminusi}
\mathcal{A}_{n-i} = 2 \sqrt{2 r_{n-i}} \sqrt{\frac{r_{n-i+1} -
r_{n-i}}{f(r_{n-i})}} + \frac{3 f(r_{n-i}) - 2 \, r_{n-i}
f'(r_{n-i})}{3 \sqrt{2 \, r_{n-i}}} \left(\frac{r_{n-i+1} -
r_{n-i}}{ f(r_{n-i})}\right)^{3/2}
\end{equation}
\begin{equation} \label{eq:ccnminusi}
\mathcal{C}_{n-i} = 2 \log{\left(2 \, r_{max}\right)} - 2
\log{\left( \sqrt{r_{n}^{2}-r_{n-i}^{2}} +
\sqrt{r_{n}^{2}+1}\right)}
\end{equation}
\begin{equation} \label{eq:bbnminusieven}
\mathcal{B}_{n-i} = \sum_{j=1}^{i/2} \frac{\left(r_{n-2j+3} -
r_{n-2j+1}\right)}{3} \left(\zeta_{n-i}(r_{n-2j+3})+ 4 \,
\zeta_{n-i}(r_{n-2j+2}) + \zeta_{n-i}(r_{n-2j+1}) \right)
\end{equation}
for $i$ even, and
\begin{equation} \label{eq:bbnminusiodd}
\mathcal{B}_{n-i} = \sum_{j=1}^{(i-1)/2} \frac{\left(r_{n-2j+2} -
r_{n-2j}\right)}{3} \left(\zeta_{n-i}(r_{n-2j+2})+ 4 \,
\zeta_{n-i}(r_{n-2j+1}) + \zeta_{n-i}(r_{n-2j}) \right)
\end{equation}
for $i$ odd, with the function $\zeta$ defined by
\begin{equation} \label{eq:newggdef1}
\zeta_{n-i}(r) \equiv \frac{1}{ \sqrt{f(r)}\sqrt{1 -
\frac{r_{n-i}^{2}}{r^{2}}}}
\end{equation}

Thus we can continue the metric extraction down to $r = 0$ in the
non-singular case, or down to $r = r_{h}$ in the black hole case
(see section \ref{sec:singspaces}).

\section*{Appendix B}

In the method of the previous appendix, the series expansion we used
to approximate part of the integral in both \eqref{eq:phir} and
\eqref{eq:lengthr} introduced an extra term, $f'(r)$, which we then
chose to linearly approximate. What appears immediately obvious is
that one could simply combine the two equations and avoid using any
approximation to $f'(r)$ at all. Considering the two terms $A_{n-i}$
and $\mathcal{A}_{n-i}$ from \eqref{eq:anminusi} and
\eqref{eq:aanminusi} respectively, we see that:

\begin{equation} \label{eq:newappendixb3}
r_{n-i} A_{n-i} - \mathcal{A}_{n-i} = \frac{4 \sqrt{2}}{3}
\frac{\left(r_{n} - r_{n-i}\right)^{3/2}}{\sqrt{r_{n-i} f(r_{n-i})}}
\end{equation}
and so by considering $r_{n-1} \, \phi_{n-1} - \mathcal{L}_{n-1}$ at
each step we eliminate the $f'(r_{n-i})$ term. For completeness, we
note that this is equivalent to the formulating the integral as
follows: beginning with expression \eqref{eq:dldphispace} and
integrating over $J$ gives:

\begin{equation} \label{eq:lphispace}
\int^{\mathcal{L}} \, \mathrm{d} \mathcal{L}' = \int J \frac{d
\phi}{d J} \, \mathrm{d} J
\end{equation}
which can then be integrated by parts:
\begin{equation} \label{eq:lphi2space}
\mathcal{L}(J) = J \, \phi(J) - \int \phi \, \mathrm{d} J
\end{equation}
and rewritten using the expression for $\phi$ from \eqref{eq:phir}:

\begin{equation} \label{eq:lphi3space}
\mathcal{L}(J) = J \, \phi(J) - \int \int_{r_{min}}^{r_{max}}
\frac{2}{r \sqrt{f(r)} \sqrt{\frac{r^{2}}{J^{2}} - 1}} \, \mathrm{d}
r \, \mathrm{d} J
\end{equation}

We can now reverse the order of integration, and as the function
$f(r)$ has no dependence on $J$, integrate over $J$. For some
specific geodesic with proper length $\mathcal{L}_{n-i}$ and angular
separation $\phi_{n-i}$ on the boundary (to continue with the
notation from earlier) we thus have that:
\begin{equation} \label{eq:lphi4space}
\mathcal{L}_{n-i} = r_{n-i} \, \phi_{n-i} + \int_{r_{n-i}}^{r_{max}}
\frac{2}{\sqrt{f(r)}} \sqrt{1 - \frac{J^{2}}{r^{2}}} \, \mathrm{d} r
\end{equation}
where we have also used that $r_{min} = J$ and relabeled the minimum
radius as $r_{n-i}$. After splitting up the integral as in Appendix
A, the lowest order approximation to the integral at $r_{n-i}$ is
given by \eqref{eq:newappendixb3}, and one can then seemingly
determine $f(r_{n-i})$, the only unknown, for each $i$ from one to
$n$ and hence reconstruct the entire metric function $f(r)$.

Applying this in practice, however, one immediately runs into the
same stability problems that occur in the naive approach mentioned
in section \ref{sec:method1}, where one attempts to recover both
$r_{min}$ and $f(r_{min})$ directly from equations \eqref{eq:phir}
and \eqref{eq:lengthr}. The method appears inherently unstable to
errors, and fails to generate any reliable estimate for $f(r)$ at
any step size. Interestingly, an almost identical formalism can be
carried out in the method involving null geodesics (see Appendix C),
however, unlike in the spacelike case, this method is both stable
and highly efficient. Further analysis into what causes the
stability/instability of the different methods is ongoing.

Finally, to clarify one further point, we note that the original
(naive) method of section \ref{sec:method1} can be stabilised by
introducing a particular regularisation of the proper length, where
one subtracts off the proper length of a corresponding geodesic in
pure AdS which probes down to the same depth, $r_{min}$. Although
this appears to not introduce any new information, one should
remember that we are working from the field theory data, and as
such, one does not in fact know the proper length of this geodesic,
but rather the one which has the same angular separation of the
endpoints. Thus using this regularisation is actually equivalent to
determining the minimum radius from \eqref{eq:dldphibspace}, using
this to determine the length of the corresponding geodesic in pure
AdS, and then treating $r_{min}$ as an unknown again in
\eqref{eq:phir5} and \eqref{eq:lengthr5}. This excessive over
complication considerably reduces the efficiency of the method, as
the equations are considerably more complicated to solve for (even
numerically) at later steps.

\section*{Appendix C}

In section \ref{sec:combo} we combine the extraction method
presented here with that given in \cite{hammer}\footnote{There are
two methods for extracting the bulk information proposed in
\protect\cite{hammer}; here we proceed to adapt the second, which is
noticeably more efficient in generating estimates for $f(r)$.} to
allow metric recovery in the most general static, spherically
symmetric spacetimes. As the methods are applied sequentially, they
require very little modification in order to work together, indeed
the spacelike method is only affected by the change in notation when
we introduce $h(r)$. The method involving null geodesics is altered
slightly more however, and so is presented in full here. This
explicit presentation also serves to highlight the similarities
between the two iterative procedures for extracting the metric,
which is remarkable given the different origins of the field theory
data.

As mentioned in the review in section \ref{sec:review}, we have a
relationship between the gradient of the endpoints of the null
geodesics (see figure \ref{modnullfig1}) and the ratio of $J$ to
$E$, namely $dt/d\phi = y$, which can be rewritten as:

\begin{equation} \label{eq:dtdp}
\frac{d t(y)}{dy} = y \, \frac{d \phi(y)}{dy}
\end{equation}
Integrating over $y$ and then by parts gives:

\begin{equation} \label{eq:tphi2}
t(y) = y \, \phi(y) - \int \phi \, \mathrm{d} y
\end{equation}
which can be rewritten by substituting in for $\phi$:

\begin{equation} \label{eq:tphi3}
t(y) = y \, \phi(y) - \int \int_{r_{min}}^{\infty} \frac{2 \, y
\sqrt{h(r)}}{r^{2} \sqrt{\frac{1}{f(r)} - \frac{y^{2}}{r^{2}}}} \,
\mathrm{d} r \, \mathrm{d} y
\end{equation}

Reversing the order of integration (as the function $f(r)$ has no
dependence on $y$) allows us to integrate over $y$:
\begin{equation} \label{eq:tphi4}
t(y) = y \, \phi(y) + \int_{r_{min}}^{\infty} 2 \sqrt{h(r)}
\sqrt{\frac{1}{f(r)} - \frac{y^{2}}{r^{2}}} \, \mathrm{d} r
\end{equation}

Thus taking the initial conditions to be $(\phi_{0}, t_{0}) =
(0,0)$, we can say that for any endpoint $(\phi_{j}, t_{j})$ on the
boundary we have:
\begin{equation} \label{eq:tphi5}
t_{j} - \frac{d t}{d \phi} \Big|_{(\phi_{j},t_{j})} \phi_{j} =
\int_{r_{j}}^{\infty} 2 \sqrt{h(r)} \sqrt{\frac{1}{f(r)} -
\frac{y_{j}^{2}}{r^{2}}} \, \mathrm{d} r
\end{equation}
where we have renamed $r_{min}$ as $r_{j}$. After using the
spacelike geodesics to determine an estimate for $h(r)$, this then
finally becomes:
\begin{equation} \label{eq:tphi5b}
t_{j} - \frac{d t}{d \phi} \Big|_{(\phi_{j},t_{j})} \phi_{j} =
\int_{r_{j}}^{\infty} 2 \sqrt{h_{\textrm{fit}}(r)}
\sqrt{\frac{1}{f(r)} - \frac{y_{j}^{2}}{r^{2}}} \, \mathrm{d} r
\end{equation}
which, when coupled with the equation for the minimum $r$,

\begin{equation} \label{eq:tphi6}
y_{j}^{2} = \frac{r_{j}^{2}}{f(r_{j})}
\end{equation}
allows the metric function $f(r)$ to be reconstructed from the plot
of the endpoints, by applying a similar iterative method to that
described in Appendix A: for the general term $r_{n-i}$, one
approximates the integral from $r_{n-i}$ to $r_{n-i+1}$ by the
parabolic area formula; the integral from $r_{n}$ to $r = \infty$ by
taking the spacetime to be pure AdS; and the remaining $i-1$
integrals by the trapezium rule, to obtain:

\begin{equation} \label{eq:tphi5c}
t_{n-i} - \frac{d t}{d \phi} \Big|_{(\phi_{n-i},t_{n-i})} \phi_{n-i}
\approx A_{n-i} + B_{n-i} + C_{n-i}
\end{equation}
where
\begin{equation} \label{eq:anminusigen}
A_{n-i} = \frac{4}{3} \, (r_{n-i+1} - r_{n-i}) \,
\eta(y_{n-i},r_{n-i+1})
\end{equation}
\begin{equation} \label{eq:bnminusigen}
B_{n-i} = \sum_{j=1}^{i-1} \left(r_{n-j+1} - r_{n-j}\right)
\left(\eta(y_{n-i},r_{n-j+1}) + \eta(y_{n-i},r_{n-j})\right)
\end{equation}
and
\begin{eqnarray}
C_{n-i} = && 2 \arctan{\left(\frac{1}{ \sqrt{\left(1 -
y_{n-i}^{2}\right)r_{n}^{2} - y_{n-i}^{2}}}\, \right)} \nonumber  \\
&& \;\; - 2 \, y_{n-i} \arctan{\left(\frac{y_{n-i}}{ \sqrt{\left(1 -
y_{n-i}^{2}\right)r_{n}^{2} - y_{n-i}^{2}}}\, \right)}
\label{eq:cnminusigen}
\end{eqnarray}
where we have defined the function $\eta(y_{j},r_{k})$ as:

\begin{equation}  \label{eq:funcdef1}
\eta(y_{j},r_{k}) \equiv \sqrt{h_{\textrm{fit}}(r_{k})}
\sqrt{\frac{1}{f(r_{k})} - \frac{y_{j}^{2}}{r_{k}^{2}}}
\end{equation}


\begin{thebibliography}{99}
%
\bibitem{thooft} G.~'t Hooft, \emph{``Dimensional reduction in quantum gravity''}, arXiv:gr-qc/9310026.
\bibitem{suss} L.~Susskind, \emph{``The World as a hologram''}, J. Math. Phys. \textbf{36}, 6377 (1995) [arXiv:hep-th/9409089].
\bibitem{adscft} J.~M.~Maldacena, \emph{``The large N limit of superconformal field theories and supergravity''}, Adv. Theor. Math. Phys. \textbf{2}, 231 (1998) [Int. J. Theor. Phys. \textbf{38},1113 (1999)][hep-th/9711200].
%
\bibitem{taka1} S.~Ryu and T.~Takayanagi, \emph{``Holographic Derivation of Entanglement Entropy from AdS/CFT''}, arXiv:hep-th/0603001.
\bibitem{taka2} S.~Ryu and T.~Takayanagi, \emph{``Aspects of Holographic Entanglement Entropy''}, arXiv:hep-th/0605073.
\bibitem{furs} D.~Fursaev, \emph{``Proof of the Holographic Formula for Entanglement Entropy''}, arXiv:hep-th/0606184.
\bibitem{solo} S.~N.~Solodukhin, \emph{``Entanglement entropy of black holes and AdS/CFT correspondence''}, Phys. Rev. Lett. \textbf{97}, 201601 (2006) [arXiv:hep-th/0606205].
\bibitem{taka3} T.~Hirata and T.~Takayanagi, \emph{``AdS/CFT and Strong SubAdditivity of Entanglement Entropy''}, JHEP \textbf{0702}, 042 (2007) [arXiv:hep-th/0608213].
\bibitem{hubtak} V.~E.~Hubeny, M.~Rangamani and T.~Takayanagi, \emph{``A Covariant Holographic Entanglement Entropy Proposal''}, arXiv:0705.0016.
%
\bibitem{hammer} J.~Hammersley, \emph{``Extracting the bulk metric from boundary information in asymptotically AdS spacetimes''}, JHEP \textbf{0612}, 047 (2006) [arXiv:hep-th/0609202].
\bibitem{vero} V.~E.~Hubeny, H.~Liu and M.~Rangamani, \emph{``Bulk-cone singularities and signatures of horizon formation in AdS/CFT''}, JHEP \textbf{0701}, 009 (2007) [arXiv:hep-th/0610041].
\bibitem{page} D.~N.~Page and K.~C.~Phillips, \emph{``Selfgravitating radiation in anti-de sitter space''}, GRG (1985).
\bibitem{radu1} D.~Astefanesi and E.~Radu, \emph{``Boson stars with negative cosmological constant''}, Nucl. Phys. \textbf{B665} (2003) 594-622 [arXiv:gr-qc/0309131].
\bibitem{radu2} D.~Astefanesi and E.~Radu, \emph{``Rotating boson stars in (2+1) dimensions''}, Phys. Lett. \textbf{B587} (2004) 7-15 [arXiv:gr-qc/0310135].
%
\bibitem{banks} T.~Banks, M.~R.~Douglas, G.~T.~Horowitz and E.~J.~Martinec, \emph{``AdS dynamics from conformal field theory''}, arXiv:hep-th/9808016.
\bibitem{mats} H.~J.~Matschull, \emph{``Black Hole creation in 2 + 1 dimensions''}, Class. Quantum Grav. \textbf{16} (1999) 1069-1095 [arXiv:gr-qc/9809087].

\end{thebibliography}
\end{document}